\documentclass[review,number,sort&compress]{elsarticle}

\usepackage{graphicx}
\usepackage{dcolumn}
\usepackage{bm}
\usepackage{relsize}
\usepackage{mdwlist}
\usepackage{amsmath}
\usepackage{amssymb}
\usepackage{color}
\usepackage{soul}
\usepackage{pifont}
\usepackage{hyperref}
\usepackage[all]{hypcap}

\def\cesrta{{C{\smaller[2]ESR}TA}}

\begin{document}

\title{Comparison of Electron Cloud Mitigating Coatings Using Retarding Field Analyzers}

\author{J.~R.~Calvey}
\author{W.~Hartung}
\author{Y.~Li}
\author{J.~A.~Livezey\fnref{berk}}
\fntext[berk]{Present address: Department of Physics, University of California, Berkeley, CA}
\author{J.~Makita\fnref{odu}}
\fntext[odu]{Present address: Department of Physics, Old Dominion University, Norfolk, VA}
\author{M.~A.~Palmer\fnref{fnal}}
\fntext[fnal]{Present address: Fermi National Accelerator Laboratory, Batavia, IL}
\author{D.~Rubin}
\address{Cornell Laboratory for Accelerator-based Sciences and
  Education, Cornell University, Ithaca, New York, USA}

\date{\today}

\begin{abstract}
    In 2008, the Cornell Electron Storage Ring (CESR) was reconfigured to serve as a test accelerator (\cesrta) for next generation lepton colliders, in particular for the ILC damping ring.   A significant part of this program has been the installation of diagnostic devices to measure and quantify the electron cloud effect, a potential limiting factor in these machines.  One such device is the Retarding Field Analyzer (RFA), which provides information on the local electron cloud density and energy distribution.  Several different styles of RFAs have been designed, tested, and deployed throughout the CESR ring.  They have been used to study the growth of the cloud in different beam conditions, and to evaluate the efficacy of different mitigation techniques.  This paper will provide an overview of RFA results obtained in a magnetic field free environment.
\end{abstract}

\maketitle


\section{\label{sec:intro} Introduction}

The electron cloud effect is a well known phenomenon in particle accelerators (see, for example,~\cite{ECLOUD12:Miguel}), in which a high density of low energy electrons builds up inside the vacuum chamber.  These electrons can cause a wide variety of undesirable effects, including emittance growth and beam instabilities~\cite{PhysRevSTAB.7.124801}.  In lepton machines, the cloud is usually seeded by photoelectrons generated by synchrotron radiation.  The collision of these electrons with the beam pipe can then produce one or more secondary electrons, depending on the secondary electron yield (SEY) of the material.  If the average SEY is greater than unity, the cloud density will grow exponentially, until a saturation is reached.

Electron cloud has been observed in many facilities (including, for example, PEP-II~\cite{PRSTAB7:024402}, CERN SPS~\cite{PRSTAB14:071001}, KEKB~\cite{CERN:SL2002:017AP}, ANL APS~\cite{PAC01:FOAB004}, FNAL Main Injector~\cite{PAC09:WE4GRC02},  LANL PSR~\cite{LANL:PSR}, and the LHC~\cite{PRSTAB16:011003}), and is expected to be a major limiting factor in next generation positron and proton storage rings.  It is of particular concern in the damping rings of electron-positron colliders, which will produce a large amount of synchrotron radiation and require very small emittances~\cite{ILCREP2007:001}.

In 2008, the Cornell Electron Storage Ring (CESR) was reconfigured to study issues related to the design of the International Linear Collider (ILC) damping ring, including electron cloud~\cite{PAC09:FR1RAI02}.  A significant component of this program, called CESR Test Accelerator (\cesrta), was the installation of several retarding field analyzers (RFAs) throughout the ring, in drift, dipole, quadrupole, and wiggler field regions.  This paper will summarize results obtained from drift RFAs.  More specifically, it will describe the the design of the detectors and experimental program (Section~\ref{sec:instrumentation}), and present measurements (Section~\ref{sec:measurements}), with a focus on directly comparing different cloud mitigation techniques.  More quantitative analysis of the RFA results will be presented in a seperate paper~\cite{ARXIV:1402.7110}.

\subsection{Retarding Field Analyzers}

\begin{figure}
	\centering
	\includegraphics[width=.45\textwidth]{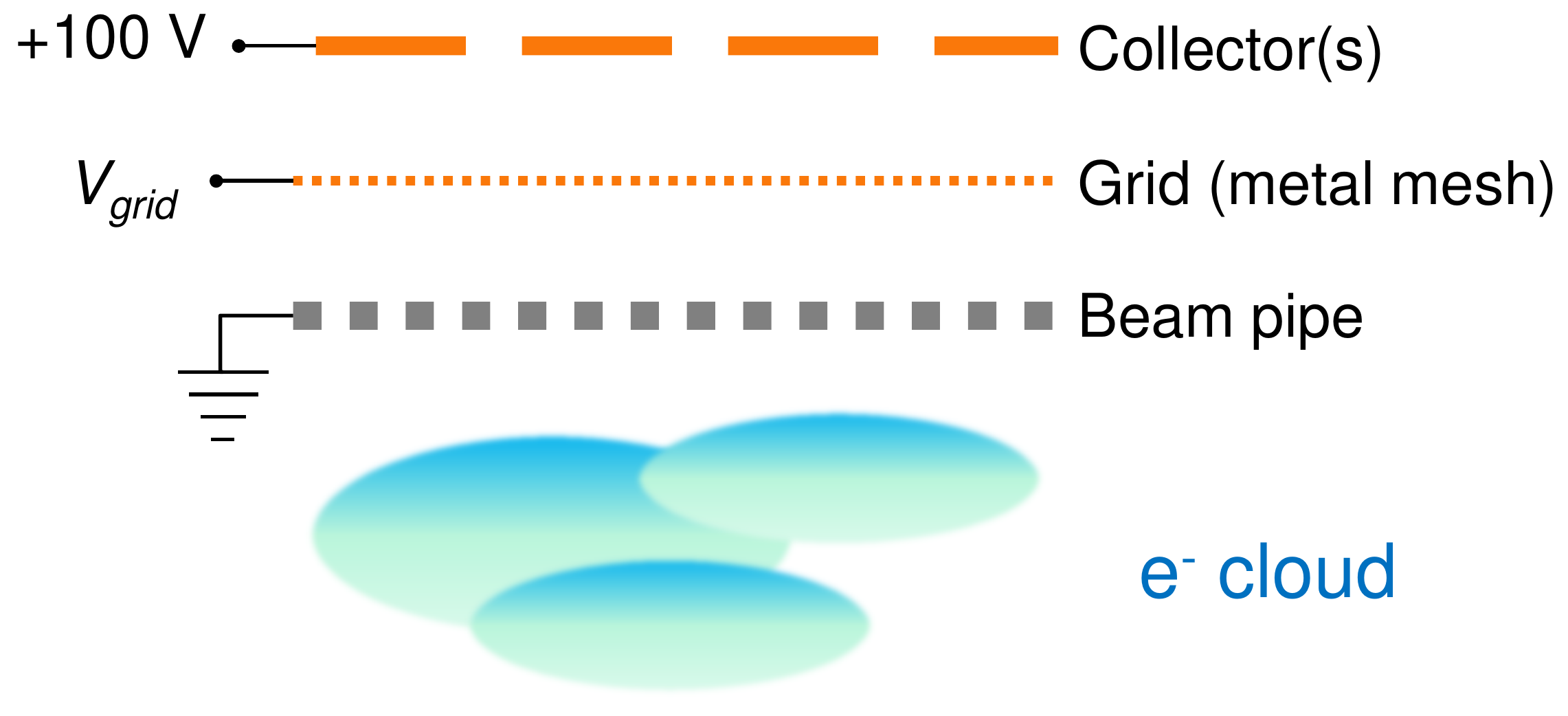}      
	\caption{Idealized diagram of a retarding field analyzer.}	
    \label{fig:rfa_diagram}
\end{figure}

A retarding field analyzer consists of three main components~\cite{NIMA453:507to513}: small holes drilled in the beam pipe to allow electrons to enter the device; a ``retarding grid," to which a voltage can be applied, rejecting electrons with less than a certain energy; and a positively biased collector, to capture any electrons which make it past the grid (Fig.~\ref{fig:rfa_diagram}).  If space permits, additional (grounded) grids can be added to allow for a more ideal retarding field.  In addition, the collectors of most RFAs used in \cesrta~are segmented transversely to allow characterization of the spatial structure of the cloud build-up.  Thus a single RFA measurement provides information on the local cloud density, energy, and transverse distribution.  Most of the data presented here are one of two types: ``voltage scans," in which the retarding voltage is varied (typically from +100 to $-250$~V or $-400$~V) while beam conditions are held constant, or ``current scans," in which the retarding grid is set to a positive voltage (typically +50~V), and data are passively collected while the beam current is increased.  When not actively in use, the RFAs were set to passively collect data, to measure the performance of the various chambers as a function of beam dose (see Section~\ref{ssec:longterm}).  The collector was set to +100~V for all of our measurements, to capture secondary electrons produced on the collector.

The use of RFAs for electron cloud studies was pioneered at APS~\cite{NIMA453:507to513}; additional studies have been performed at the FNAL Main Injector~\cite{PAC09:TH5RFP041}, PEP-II~\cite{NIMA621:33to38}, and KEKB~\cite{PAC05:FPAP007}.  However, the \cesrta~RFA program is unprecedented in terms of scale.  We have used RFAs to probe the local behavior of the cloud at multiple locations in CESR, under many different beam conditions, and in the presence of several different mitigation schemes.


A few additional considerations were important in the design of the CESR RFAs:

\begin{itemize}
    \item Some designs needed to fit into confined spaces ($\sim$2-3~mm), such as the aperture of the CESR dipole magnets.
    \item The detectors needed to be shielded from direct beam signal.  A 3:1 depth to diameter ratio for the beam pipe holes was determined to be sufficient to effectively shield the RFAs.
    \item Production of secondary electrons inside the detector should be minimized.  To accomplish this, most of the grids were coated with gold, which has a low secondary electron yield.
\end{itemize}

\subsection{Experimental Sections}

\begin{figure}
	\centering
	\includegraphics[width=0.45\textwidth]{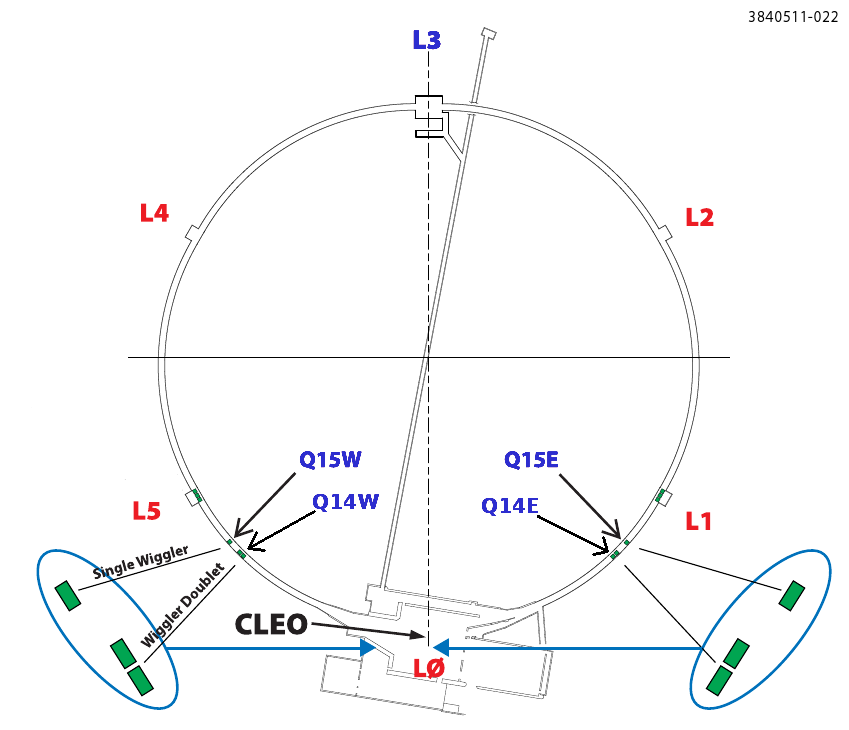}     
	\caption[{\cesrta} Vacuum System]{The reconfiguration of the CESR vacuum system provided space for several electron cloud experimental sections.  Drift RFAs are located at Q14E/W, Q15E/W, and L3. \label{fig:cesr_configuration}}

\end{figure}

There are five main electron cloud experimental sections of CESR instrumented with drift RFAs.  These include long sections at Q14E and Q14W (the names refer to their proximity to the 14E and 14W quadrupoles, respectively), shorter sections at Q15E and Q15W, and a long straight section at L3.  The vacuum chambers at Q15E/W are approximately elliptical and made of aluminum (6063 alloy, as is most of CESR); the chambers at Q14E/W are approximately rectangular and made of copper; the pipe at L3 is circular and stainless steel.  The specific needs of each experimental section necessitated the design of several different types of drift RFA (Section~\ref{ssec:styles}). Fig.~\ref{fig:cesr_configuration} shows the locations of these experimental sections in the CESR ring; more details on each location are given in Section~\ref{ssec:sections}.

\subsection{CESR Parameters}

The primary advantage of CESR as a test accelerator is its flexibility.  At \cesrta, we have been able to study the behavior of the electron cloud as a function of several different beam parameters, a small subset of which are presented here (additional measurements can be found in ~\cite{PHD:Me}).  Table~\ref{tab:mitigation_beam_condx} gives some of the basic parameters of CESR, and lists some of the beam parameters used for electron cloud mitigation studies with RFAs.

\begin{table}
   \centering
   \caption{\label{tab:mitigation_beam_condx} CESR parameters and typical beam conditions for electron cloud mitigation studies.}
   \begin{tabular}{ccc}
       \hline \hline
       Parameter &   Value(s)  &   Units \\
       \hline
            Circumference   &   768 &   m   \\
            Revolution Period   &   2.56    &   $\mu$s \\
            Harmonic number     &   1281    &   -   \\
            Number of bunches  &   9, 20, 45   &   -  \\
            Bunch spacing  &   4 - 280   &   ns   \\
            Beam energy &   5.3  &   GeV  \\
            RMS Horizontal Emittance    &   144     &   nm  \\
            RMS Vertical Emittance  &   1.3    &   nm   \\
            RMS Bunch Length    &   20.1  &   mm  \\
            Bunch current  &   0 - 10   &   mA   \\
            Beam species &   e$^+$, e$^-$  &   - \\
        \hline \hline
   \end{tabular}
\end{table}

\subsection{Cloud Mitigation}

In addition to solenoid windings (which trap electrons near the vacuum chamber wall~\cite{PRSTAB7:024402}), the primary method of reducing electron cloud density in a field free region is the use of beam pipe coatings, which reduce the primary and/or secondary emission yield of the chamber.  Coatings tested at \cesrta~include titanium nitride (TiN)~\cite{NIMA551:187to199}, amorphous carbon (aC)~\cite{PRSTAB14:071001}, diamond-like carbon (DLC)~\cite{ECLOUD10:MIT00}, and Ti-Zr-V non-evaporable getter (NEG)~\cite{NIMA554:92to113}.  More details on the various coated chambers have been published elsewhere~\cite{IPAC13:THPFI088}.

\section{\label{sec:instrumentation} Instrumentation}

The design of the RFAs has evolved over the course of the \cesrta~program since it began in mid 2008.  A thorough account of the design and construction of the RFAs can be found in~\cite{PHASE1_REPORT_VACUUM}; here we provide an overview.

\subsection{\label{ssec:styles} RFA Styles}

Several different styles of RFA have been deployed throughout drift sections in CESR.  Table~\ref{tab:rfa_styles} summarizes the key parameters of each style, and Table~\ref{tab:grid_styles} describes the different types of grids used.  A more detailed description of each RFA style follows:

\begin{table}
   \centering
   \caption{\label{tab:rfa_styles} Drift RFA styles deployed in CESR.  Each RFA has one retarding grid.  For RFAs with multiple grids, the additional grids are grounded.}
   \begin{tabular}{cccc}
       \hline \hline
       Type &   Grids   &   Collectors  &   Grid Type\\
       \hline
            Thin Test   &   1   &   1   &   Etched   \\
           APS  &   2   &   1   &    Mesh \\
           Insertable I   &   2   &   5  &  Etched   \\
           Insertable II  &   3   &   11   &   HT Mesh   \\
           Thin  &   1   &   9   &  HT Mesh   \\
        \hline \hline
   \end{tabular}
\end{table}

\begin{table}
   \centering
   \small
   \caption{\label{tab:grid_styles} Grid types used in CESR RFAs.  Note that ``transparency" refers to the optical transparency.}
   \begin{tabular}{cccc}
       \hline \hline
       Type &    Transparency   &   Material  &   Thickness \\
       \hline
           Etched   &   ~40\%   &   Gold coated SST   &   150~$\mu$m   \\
           Mesh  &   ~46\%   &   SST   &   76~$\mu$m \\
           HT Mesh   &   ~90\%   &   Gold coated copper  &  13~$\mu$m   \\
        \hline \hline
   \end{tabular}
\end{table}

\paragraph{APS style} This design is based on a well understood style of RFA, originally used at APS~\cite{NIMA453:507to513}.  It consists of a single collector, and two stainless steel meshes for grids.  APS style RFAs were deployed at Q14E, as well as the L3 NEG test chamber (Section~\ref{ssec:NEG}).

\begin{figure}
	\centering
	\includegraphics[width=.45\textwidth]{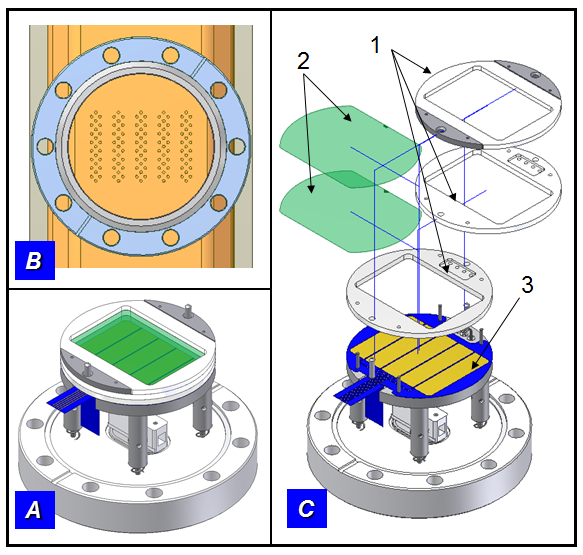}
	\caption{Engineering diagram of an ``Insertable I" style RFA. (A) Assembled RFA structure.  (B) Vacuum chamber hole pattern.  (C) Exploded view of the RFA, showing (1) Macor spacers, (2) stainless steel etched grids, and (3) flexible circuit collectors.}
    \label{fig:insertable_segmented_rfa}
\end{figure}

\paragraph{Insertable Type I} Deployed in the Q14E and Q14W experimental regions of CESR, these RFAs were designed to be ``inserted" in a port on top of a specialized vacuum chamber.  They have two grids, spaced by 3~mm.  The grids are stainless steel (SST), with an etched bi-conical hole structure (0.18~mm diameter holes with a 0.25~mm pitch).  In addition, the grid layer was vacuum-coated with a thin gold layer (several hundred~nm) to reduce its secondary electron yield.  The electron collector pad was laid out on copper-clad Kapton sheet using standard printed circuit board fabrication techniques.  Transverse resolution is obtained by using five transversely arranged collectors.  Holes are drilled in the beam pipe in five segments; each segment has 25 holes, with diameter 1.5~mm and depth 5.1~mm.  Fig.~\ref{fig:insertable_segmented_rfa} gives a detailed picture of this RFA.

\begin{figure}
	\centering
	\includegraphics[width=.45\textwidth]{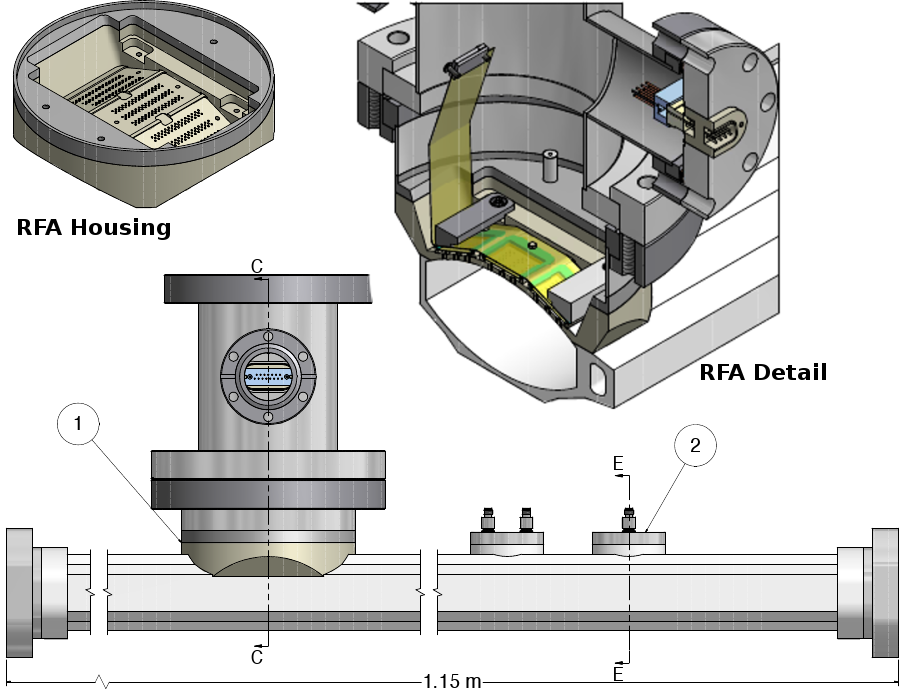}  
	\caption{Q15 EC Test Chamber, equipped with an RFA (1) and 4 shielded pickups (2). \label{fig:Q15_vc}}	
\end{figure}

\paragraph{Thin style} Designed for use inside a CESR dipole, where vertical aperture space is limited, the thin style detector was also used in the Q15E and Q15W drift sections (which use the same elliptical beam pipe extrusion as the dipole).  The RFA housing is machined from a separate block of explosion-bonded aluminum-to-stainless steel material, and is welded to the cutout on top of the beam pipe.  The lower face of the RFA housing matches the curvature of the beam pipe aperture, while the upper face is divided into three flat sections.  Each section has one retarding grid, which is made of high efficiency electro-formed high transparency (HT) copper mesh, held in place by a stainless steel frame.  There are three collectors in each section, for a total of nine.  The total distance from the outside of the vacuum chamber to the collectors is 2.5~mm.  The beam pipe holes are 0.75~mm in diameter and $\sim$2.5~mm in thickness, maintaining the same ratio of diameter to thickness used for the ``Insertable I" style.  There are 44 holes per collector.  A diagram of a Q15 test chamber, which includes a thin RFA (modified for use in a drift space), as well as 4 shielded pickup detectors~\cite{ARXIV:1311.7103}, is shown in Fig.~\ref{fig:Q15_vc}.

\paragraph{Insertable Type II}    The second generation insertable RFA has three grids, consisting of high-transparency copper meshes, spaced by 5.7~mm. The retarding voltage is applied to the middle grid.  Insertable II RFAs were installed in the 5$^{th}$ Q15 test chambers, replacing the ``thin" style detectors.  The additional grids and increased spacing between them allows for higher retarding voltage (up to $-400$~V), and reduces some of the non-ideal behavior observed in the thin RFAs (Section~\ref{ssec:bench}).  To provide cross calibration between the two RFA designs, a TiN-coated test chamber in Q15W was instrumented with both styles.  The two styles gave similar results.

\subsection{\label{ssec:sections} Experimental Sections}

Table~\ref{tab:rfa_master_list} summarizes the location of each drift RFA.  The purpose of each experimental section is detailed below.

\begin{table}
  \centering
  \scriptsize
    \caption{\label{tab:rfa_master_list} List of drift RFA locations.  ``Material" refers to the base material; some locations have tested one or more coatings.  The vacuum chambers at all locations are 5~cm in height by 9~cm in width, with the exception of the circular chambers, which are 4.5~cm in radius.}
  \begin{tabular}{ccccc}
    \hline \hline
  Location  &   RFA Type    &   Material    &   Coatings  &     Shape \\
  \hline
  14W   &   Ins.   &   Cu  &  TiN   &   Rectangular \\
  15W   &   Thin, Ins. II   &   Al  &   TiN, aC &   Elliptical  \\
  L3    &   APS   &   SST &   NEG   &   Circular    \\
  15E   &   Thin, Ins. II   &   Al  &   TiN, aC, DLC    &   Elliptical  \\
  14E   &   APS, Ins.   &   Cu  &  TiN   &   Rectangular \\
  \hline \hline
  \end{tabular}
\end{table}

\subsubsection{\label{ssec:14EW} Q14W and Q14E Test Sections} Upon the removal of the CESR-c superconducting wigglers~\cite{PAC09:TH5RFP029}, two electron cloud experimental sections were created in both the east and west arcs of CESR.  Measurements in the Q14W test section confirmed that an ``Insertable Type I" style RFA gives results comparable to the well validated ``APS" style~\cite{PAC09:TH5RFP030}.  At Q14E, the copper beam pipe was coated with TiN thin film for half of its length (while the other half remained bare copper). Insertable RFAs were installed at each end of this test chamber to compare electron cloud intensity in the two sections.

\subsubsection{\label{ssec:15EW} Q15W and Q15E Mitigation Comparison Chambers} 
\begin{figure}
	\centering
	\includegraphics[width=0.45\textwidth]{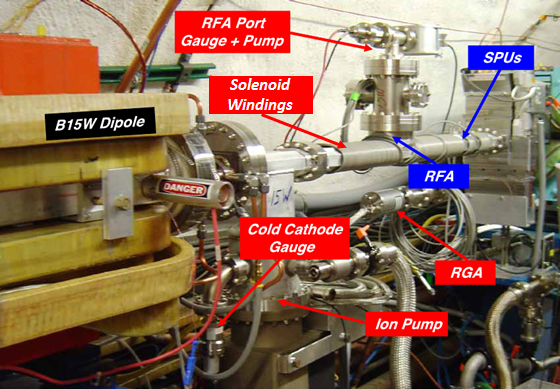}     
	\caption{A Q15 experimental chamber installed at Q15W in CESR.  In addition to the RFA, the chamber contains 4 shielded pickups (SPUs)~\cite{ARXIV:1311.7103}. \label{fig:cesr_conversion:Q15W_installed}}	
\end{figure}

To allow for frequent exchange of the test chambers while minimizing the impact on the accelerator operations, two very short ($\sim$1~m) experimental regions were created in the Q15W and Q15E locations in the arcs.  Over the course of the \cesrta~program, four chamber surfaces were tested in these locations: bare aluminum (as it was originally extruded), aC coatings (coated by CERN/CLIC), TiN coating (by Cornell) and DLC coating (by KEK).  Table~\ref{tab:Q15_chamber_table} gives detailed information on these chambers, and Fig.~\ref{fig:cesr_conversion:Q15W_installed} shows a typical installation at Q15W.

\begin{table*}

\begin{minipage}{\textwidth}
\centering
\setlength{\tabcolsep}{6pt}
\scriptsize
\caption{Summary of Q15W and Q15E Experimental Vacuum Chambers (VCs). \label{tab:Q15_chamber_table}}
\begin{tabular}{lcccccp{0.275\textwidth}}
\hline \hline
VC & Surface & Run & RFA Style & Test Period & Location & Note \\ \hline
1 & Al  & 1 & Thin & Jul 2009-Nov 2009 & E & Reference surface \\
  &     & 2 & Thin & Apr 2010-Aug 2010 & W & \\
  &     & 3 & Insertable II & Aug 2012-present & E & \\ \hline
2 & TiN & 1 & Thin & Dec 2009-Apr 2010 & E & Coated by DC sputtering at Cornell \\
  &     & 2 & Thin & Aug 2010-Jan 2011 & W & \\
  &     & 3 & Insertable II & Feb 2011-Jul 2011 & W & \\
  &     & 4 & Insertable II & Aug 2012-present & W & \\ \hline
3 & aC  & 1 & Thin & Jul 2009-Apr 2010 & W & Coated by DC sputtering at CERN\\ \hline
4 & aC  & 1 & Thin & Apr 2010-Jan 2011 & E & Coated by DC sputtering at CERN\\
  &     & 2 & Insertable II & Jul 2011-Jul 2012 & W & \\ \hline
5 & DLC & 1 & Insertable II & Feb 2011-Jul 2012 & E & Coated by pulsed DC chemical vapor deposition, supplied by KEK\\
\hline \hline
\end{tabular}
\end{minipage}
\end{table*}

There is some evidence that the aC coated chambers may have been contaminated by silicone tape present during the bakeout of the chamber~\cite{PHASE1_REPORT_VACUUM}, raising the effective SEY.  However, as described in Section~\ref{sec:measurements}, these chambers still showed good performance in situ.

\subsubsection{\label{ssec:NEG} L3 Test Section} 
\begin{figure*}
    \begin{minipage}{\textwidth}
    	\centering
	\includegraphics[width=\textwidth]{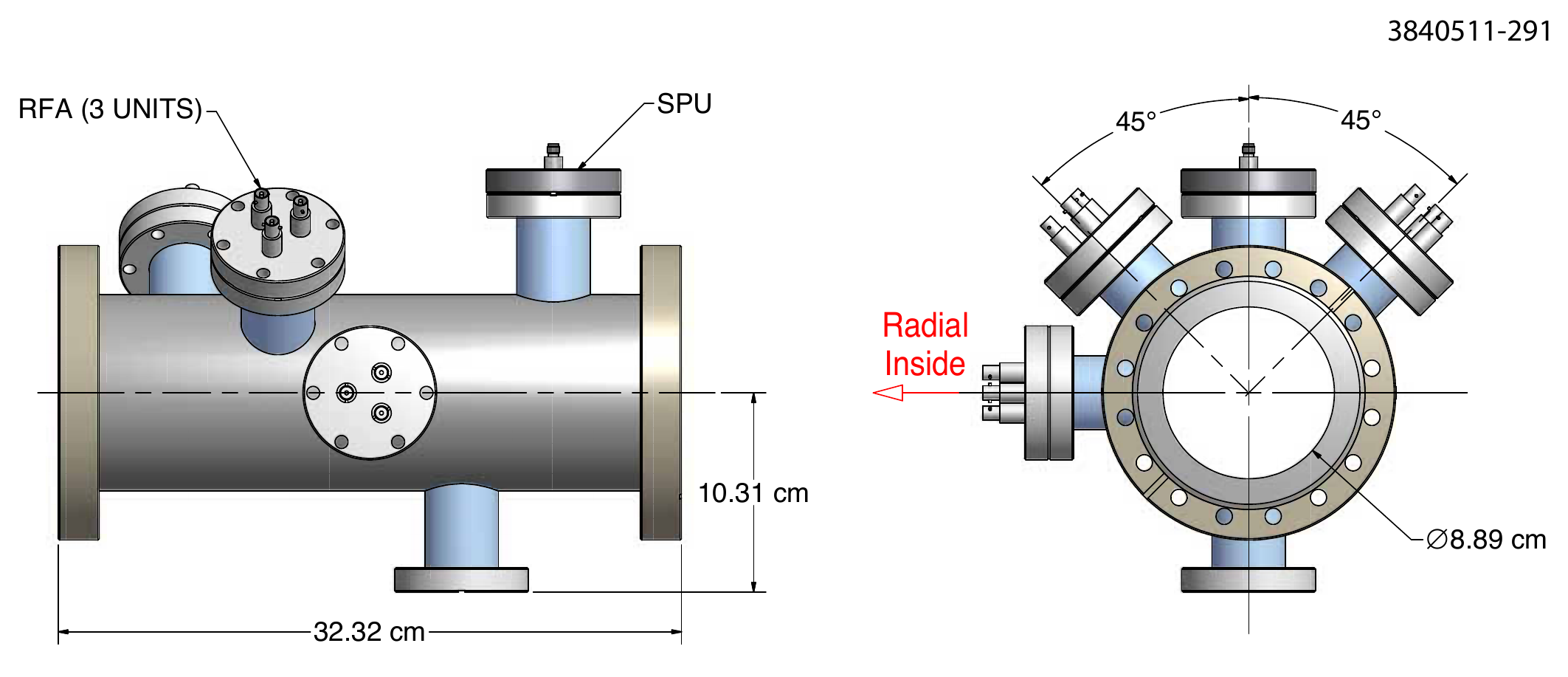}
	\caption[NEG coated EC Chamber in L3]{Electron cloud diagnostic chamber with NEG thin film coating. \label{fig:cesr_conversion:NEG_pipe_diagnostics}}	
    \end{minipage}
\end{figure*}

A Ti-Zr-V non-evaporable getter (NEG) thin film \cite{NIMA554:92to113} has been shown to have a low SEY, after its activation at elevated temperatures under vacuum.  The activated NEG coating also has the benefit of providing vacuum pumping.  A NEG-coated test chamber was built and tested in the drift section of the L3 experimental region in CESR.  To prevent rapid saturation of the activated coating from residual gases in the surrounding beam pipes, the test chamber was sandwiched between two 1-m long NEG coated beam pipes.  The chamber was equipped with three APS-style RFAs at three different azimuthal angles (see Fig.~\ref{fig:cesr_conversion:NEG_pipe_diagnostics}).  All three chambers were made of stainless steel (Type 304L).

\subsection{\label{ssec:daq} Data Acquisition}

A modular high voltage power supply and precision current monitoring system were designed to support the RFA measurements.  Each experimental section is controlled by an electronics crate with up to three high voltage power supplies (with each supply corresponding to a single style of RFA).  The individual RFAs are controlled and monitored by data boards installed in the crate.  The crates are located in the tunnel to reduce the length of cables.  Each crate is connected to the CESR control system through the local fieldbus.  More details on the RFA electronics can be found in~\ref{sec:electronics}.


We used coaxial cables for the single-collector RFAs.  To reduce the volume of cables, we originally used a single multi-wire foil-shielded cable for each segmented-collector RFA.  In early tests, we observed anomalously long-time-scale transient currents.  This problem was remedied by using three separately-shielded cables, one for the collectors (biased at $+100$~V), one for the retarding grid (with varying bias), and one for the grounded grid(s).  To mitigate possible leakage currents to ground in the cables, the outer shields were biased to the same voltage as the signal wires.

Data acquisition code, capable of both voltage scans and continuous monitoring of collector current, is interfaced through the CESR control system.  The data acquisition is controlled through a MATLAB\footnote{http://www.mathworks.com/products/matlab/} based graphical user interface (GUI), which also allows for real time monitoring and control of any RFA.  Data are collected simultaneously for all the RFAs, and the GUI allows for commands to be issued to all the devices at once.  The GUI can also be used to load predefined ranges (see~\ref{sec:electronics}) for each channel or to adjust the range for each channel based on the measured current.


While commissioning the RFA electronics with no beam in CESR, we observed a slowly varying signal, on the order of ~.2\% full scale (corresponding .02 - 20 nA, depending on the range setting).  To mitigate the effect of this signal on the experiments, a baseline measurement is always taken before a set of RFA data, and this baseline current is subtracted from all subsequent measurements.

Collecting a single data point with all RFAs takes a few seconds.  Typically, voltage scans are done with 20~V steps, with three data points per step.  After each voltage step, the DAQ software pauses to avoid recording transient currents, and then collects data.  A full voltage scan with all RFAs takes approximately 6 minutes.

\subsection{\label{ssec:bench} Bench Tests}

\begin{figure}
\centering
\begin{tabular}{c}
\includegraphics[width=.45\textwidth]{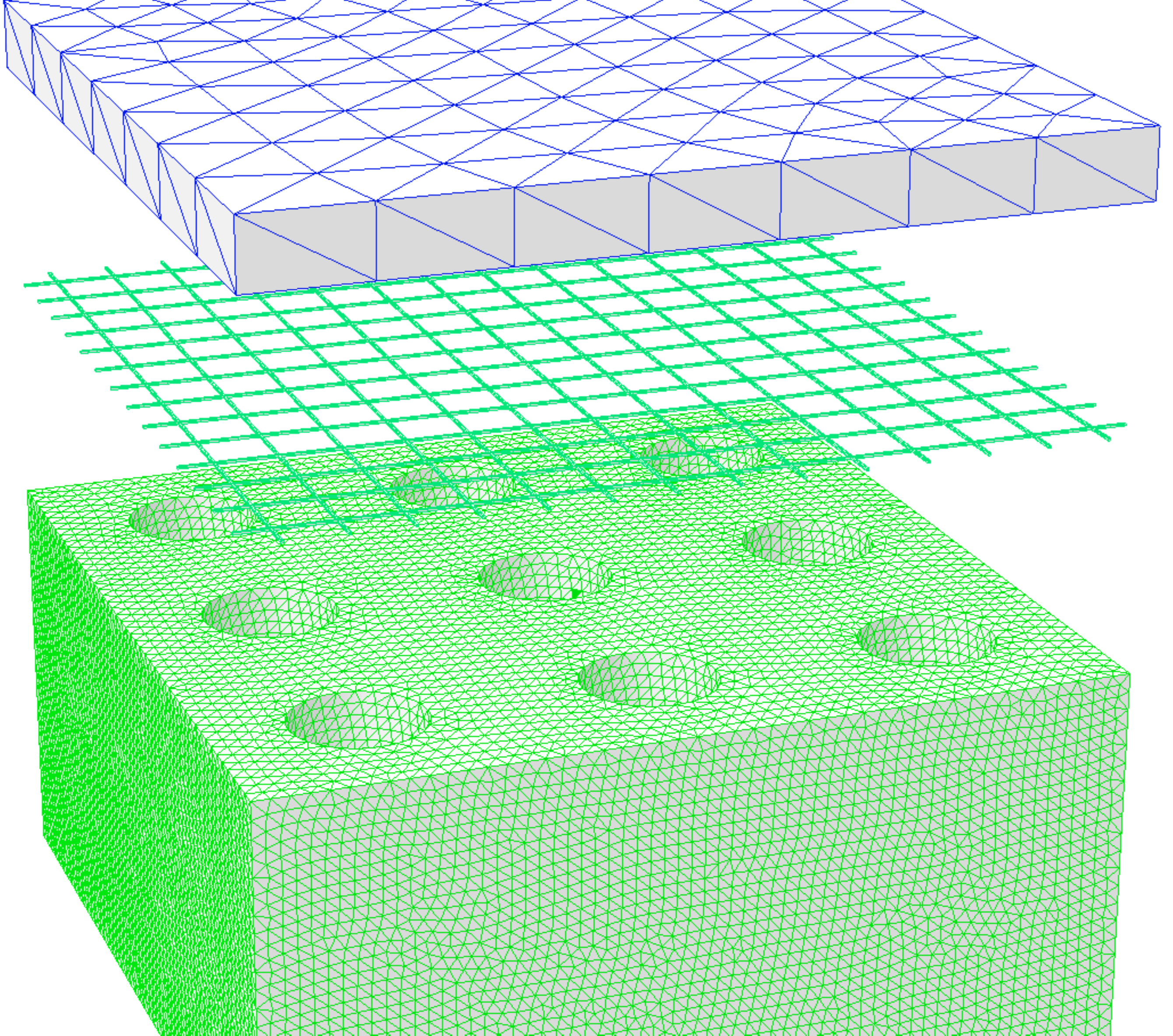} \\
\includegraphics[width=.45\textwidth]{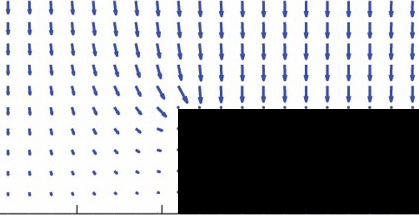} \\   
\end{tabular}
\caption{\label{fig:model} Opera 3D model of a thin RFA.  Top: full model, showing (from top to bottom) the collector, thin retarding grid, and faceplate/vacuum chamber.  Bottom: calculated electric field (blue arrows) near a hole in the faceplate (black rectangle).}
\end{figure}

Initial bench tests, which verified reasonable operation of our RFAs, have been published previously~\cite{PAC09:TH5RFP030}.  More recently, we constructed a more sophisticated bench experiment to study the response of a test RFA under controlled conditions.  We also developed a specialized particle tracking code, which tracks electrons through a model of the RFA.  The model includes a detailed replica of the beam pipe, grid(s), and collector, as well as a realistic map of the electric fields inside the RFA, generated by the electrostatic calculation tool Opera 3D\footnote{http://operafea.com/} (Fig.~\ref{fig:model}).  The tracking code also allows for the production of secondary electrons on both the beam pipe and grid(s).  More information on the tracking model has been published elsewhere~\cite{PAC11:MOP214,ARXIV:1402.7110}.


\begin{figure}
\centering
\begin{tabular}{cc}
\includegraphics[width=.45\linewidth]{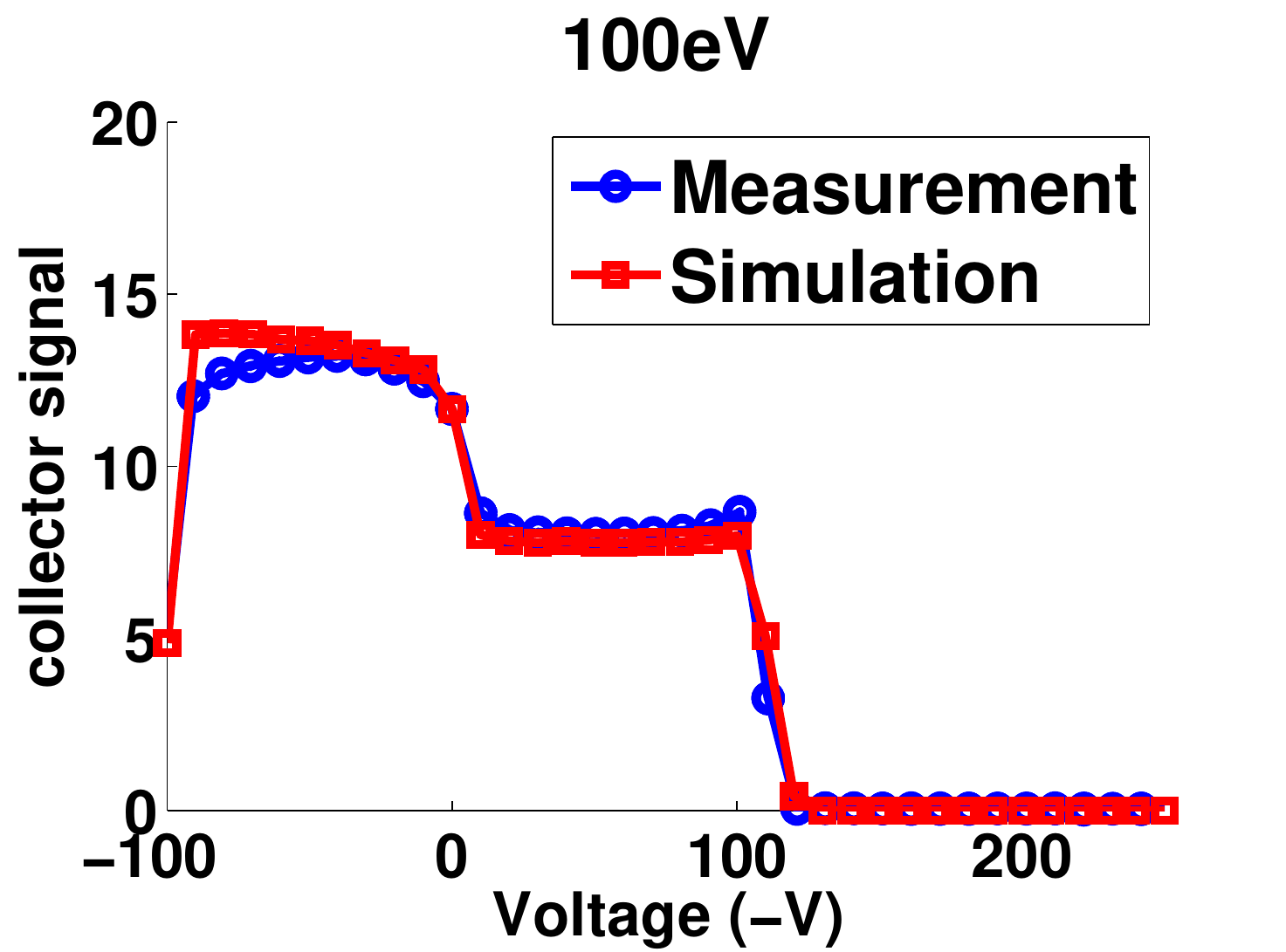}  
\includegraphics[width=.45\linewidth]{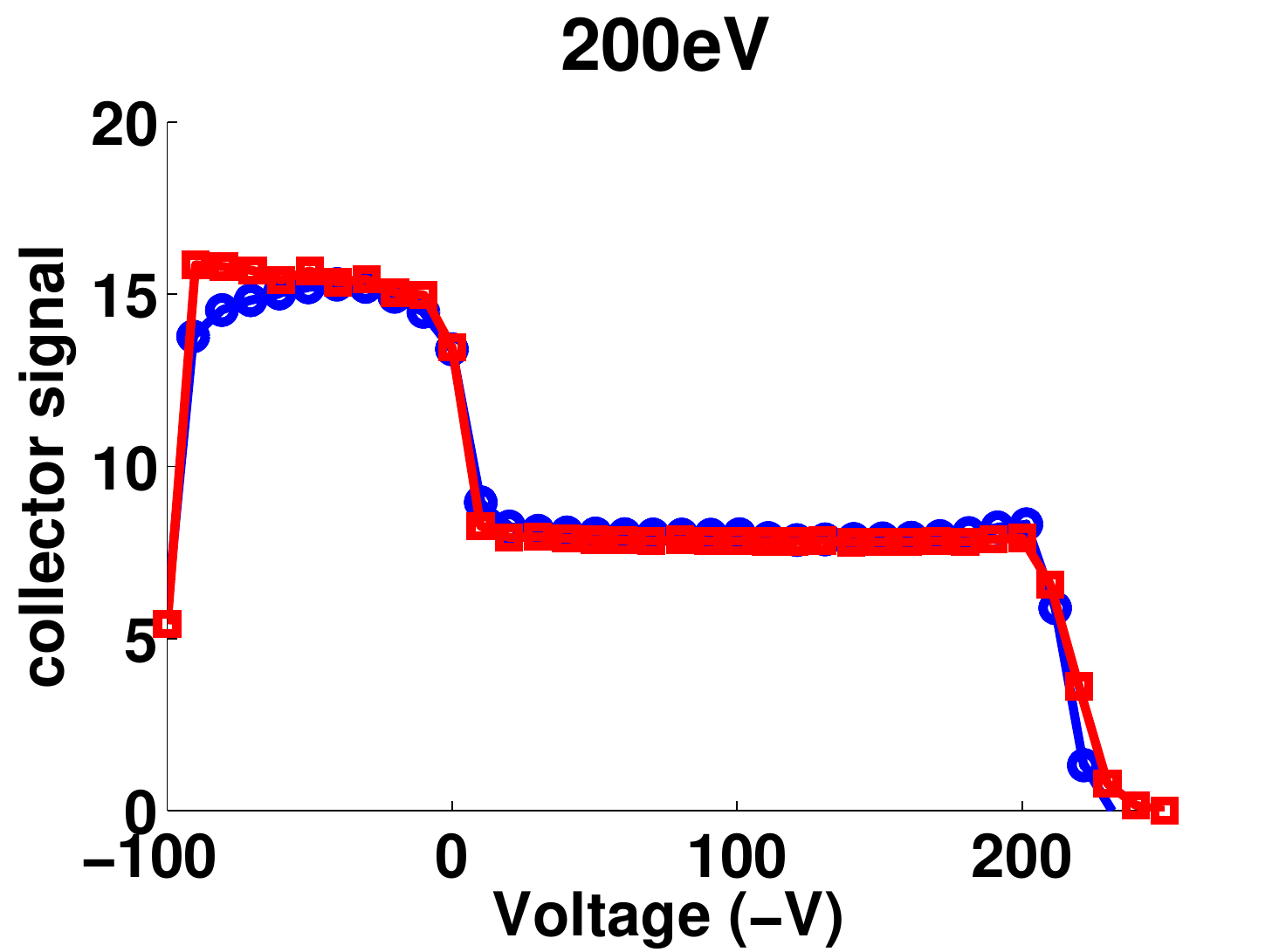} \\ 
\includegraphics[width=.45\linewidth]{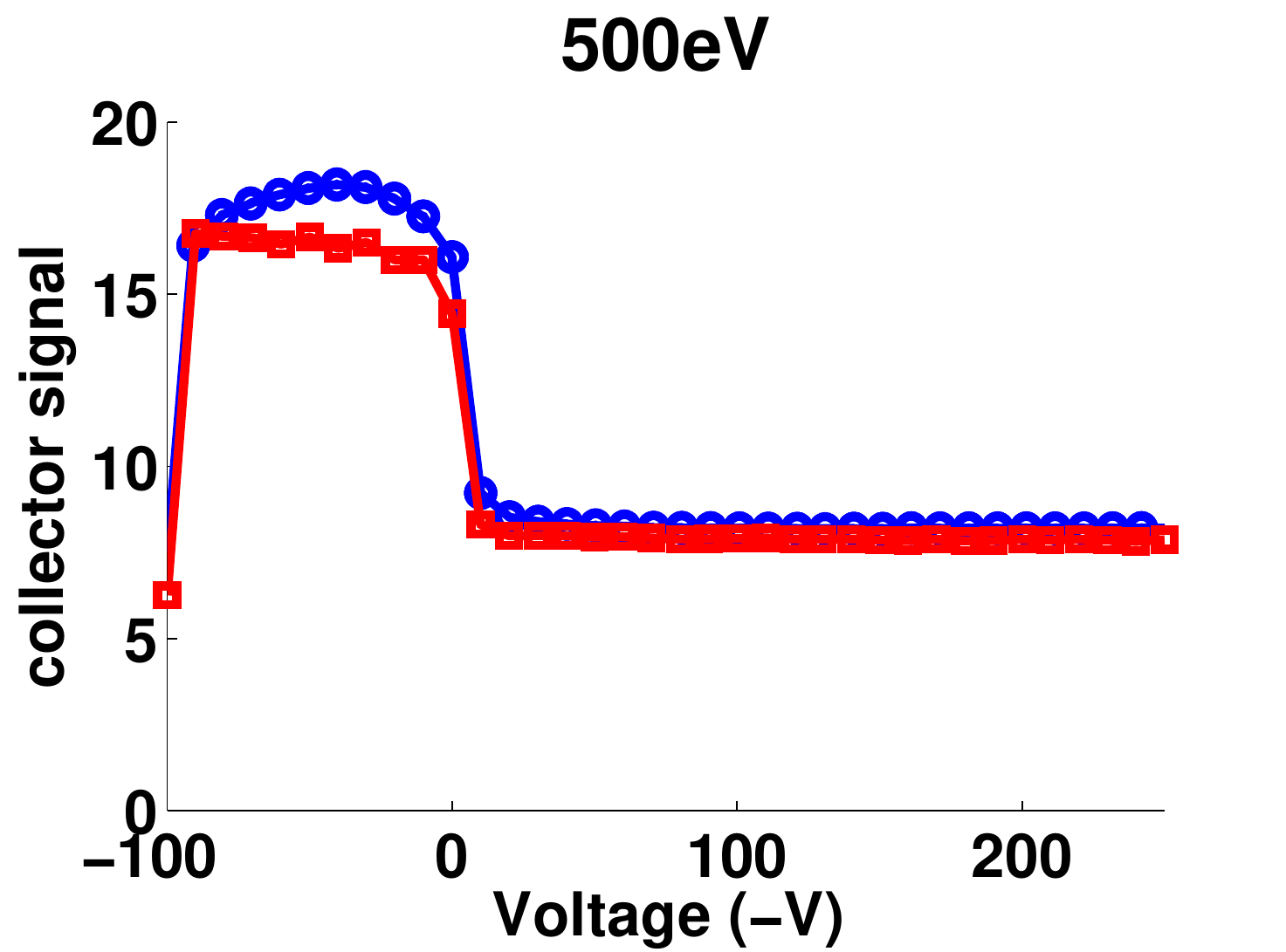}  
\includegraphics[width=.45\linewidth]{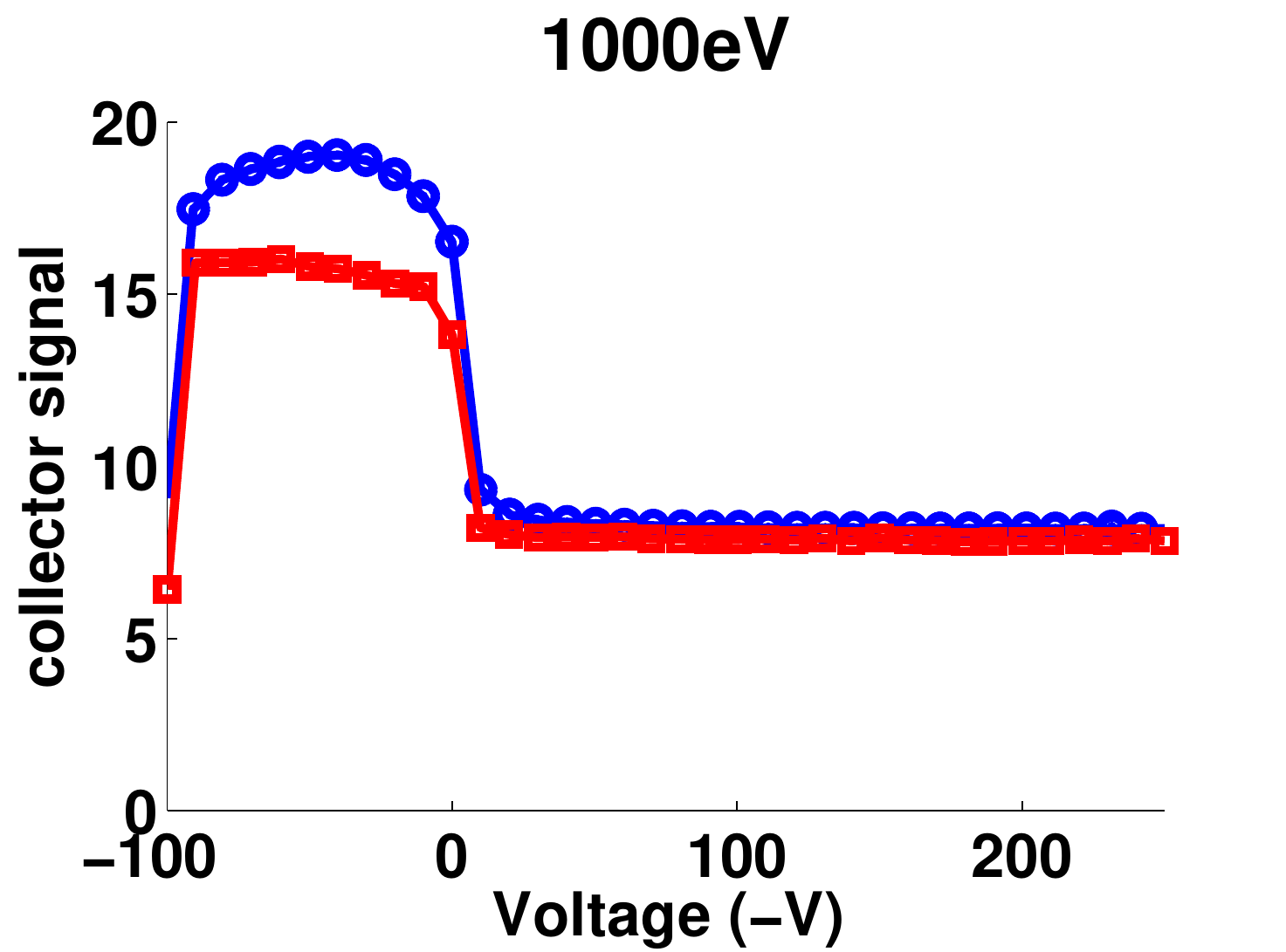} \\ 
\end{tabular}
\caption{\label{fig:bench_measurements} Comparison of bench measurement and simulation, with electron gun energy 100~eV (top left), 200~eV (top right), 500~eV (bottom left), and 1~keV (bottom right).}
\end{figure}

The bench setup consists of a DC electron gun\footnote{Model ELG-2, Kimball Physics, Inc., Wilton, NH.}, which can produce a monoenergetic and roughly uniform beam of electrons, aimed at a test RFA (Fig.~\ref{fig:bench_diagram}).  The electron gun and RFA are installed in a vacuum chamber with mu metal for shielding of ambient magnetic fields.  The RFA includes a faceplate with holes drilled in it to mimic the vacuum chamber, a high efficiency (nominally $92\%$) retarding grid, and a collector.  This mimics the structure of the ``thin style" RFA (Table~\ref{tab:rfa_styles}, except that the test RFA has only one collector).  We are able to independently control the voltage and read the current on the collector, grid, and faceplate, as well as a separately grounded top ring surrounding the faceplate.  To do a measurement with this system, we set the electron gun to a specific energy, and adjust the deflection and focusing of the gun until the beam just covers the faceplate (i.e. until no current is observed on the top ring).  We can then study the response of the RFA as a function of gun energy.   Fig.~\ref{fig:bench_measurements} shows a series of retarding voltage scans done with our bench setup at different electron gun energies, and compares them to predictions from the particle tracking model (assuming a uniform angular distribution from the electron gun).  A few things are worth noting about these measurements:

\begin{figure}
\centering
\includegraphics[width=.45\textwidth]{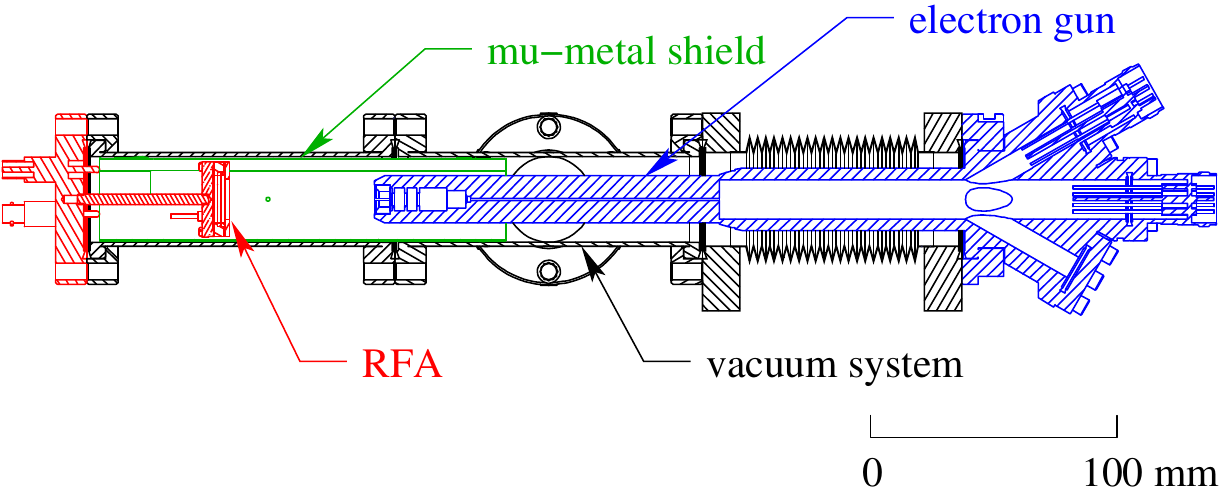}   
\caption{\label{fig:bench_diagram} Diagram of RFA bench test setup.}
\end{figure}


\begin{itemize*}
    \item The collector signal is mostly flat for a retarding voltage between 0 and the gun energy, as expected for a monoenergetic beam.
    \item When the grid voltage is positive, there is a strong enhancement of the signal, caused by the production of low energy secondary electrons in the faceplate holes.
    \item With +100~V on the grid (on the left side of the plots), the signal drops back down somewhat.  This is because secondaries produced on the collector (which is also set to +100~V) are now able to escape.
    \item If the RFAs were ideal, the collector signal would drop to zero when the retarding voltage exceeds the gun energy.  In the 100~eV and 200~eV scans, the signal does not immediately vanish, but drops off steadily, reaching zero current at $-120$~V and $-230$~V respectively. This effect is caused by focusing of the electrons by the non-ideal field of the grid, which allows electrons with energy slightly lower than the retarding voltage to slip by.  This effect has also been observed in studies of RFA performance done at FNAL~\cite{PAC09:TH5RFP041}.
\end{itemize*}


The simulation matches all important features of the data, including the enhancement at positive voltage and the non-ideal energy cutoff.  The agreement is nearly perfect for 100~eV, and 200~eV, but the simulation slightly underestimates the collector signal at positive voltage for 500~eV and 1~keV.  This aspect of the data is not understood, but could be due to a change in the beam profile at high gun energy, which is not included in our model.  Nonetheless, the agreement between the measurement and model is excellent overall.  These tests verified that we understand basic operation of our RFAs, and were used as input for detailed simulations~\cite{ARXIV:1402.7110}.

\section{\label{sec:measurements} Measurements}

\begin{figure}
   \centering
   \includegraphics*[width=0.5\textwidth]{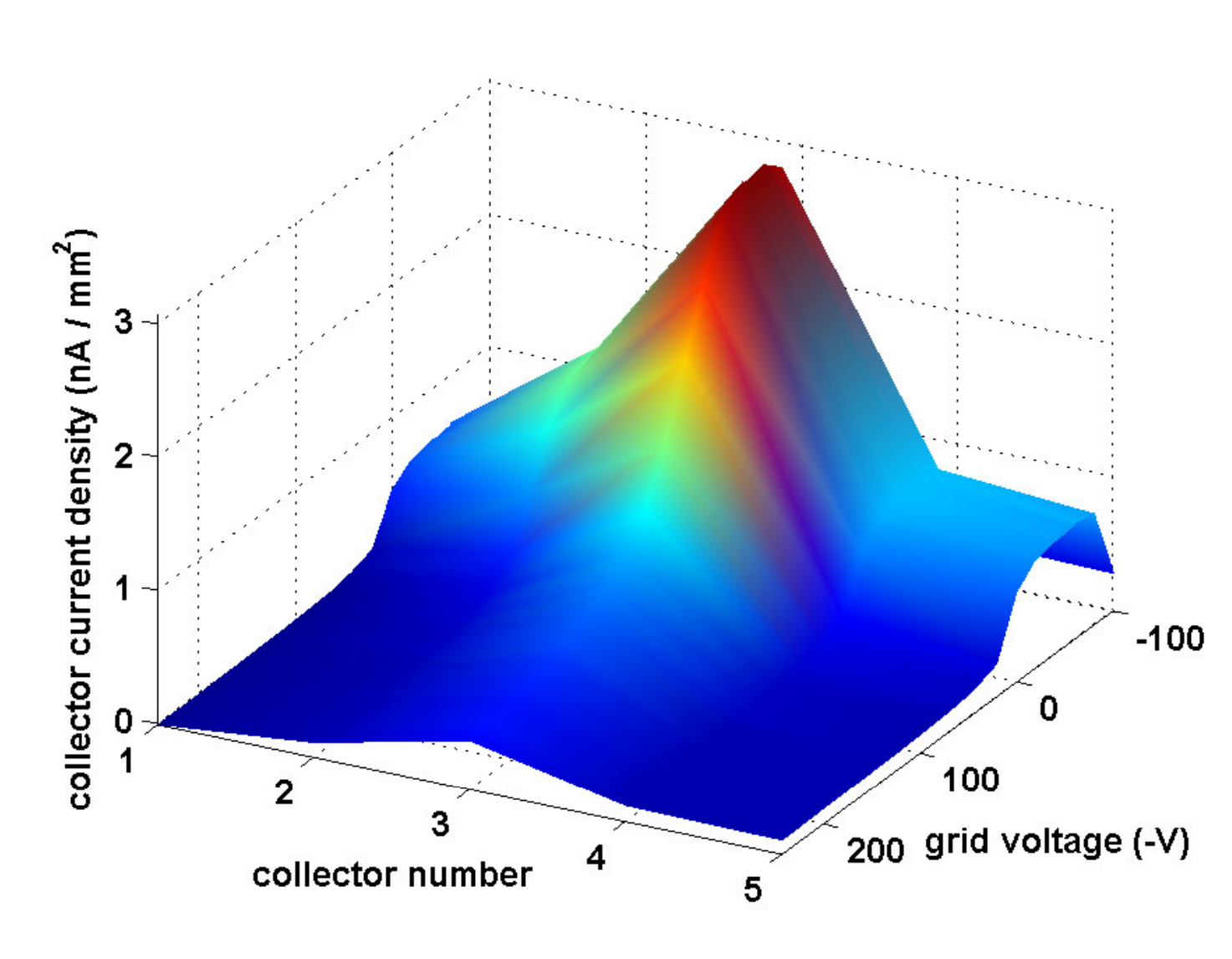}
   \caption[RFA voltage scan with an insertable segmented drift RFA]{\label{fig:seg_example} RFA voltage scan with an ``Insertable Type I" style drift RFA in a Cu chamber, 1x45x1.25 mA e$^+$, 14~ns, 5.3 GeV.}
\end{figure}

Many of our earliest detailed measurements were done with ``Insertable Type I" style RFAs (Table~\ref{tab:rfa_styles}).  Fig~\ref{fig:seg_example} shows an example of a voltage scan done with one of these detectors, in typical \cesrta~beam conditions.  The RFA response is plotted as a function of collector number and retarding voltage.  Roughly speaking, this is a description of the transverse and energy distribution of the cloud.  Collector  1 is closest to the outside of the chamber (where direct synchrotron radiation hits); the central collector (3 in this case) is aligned with the beam.  The sign convention for retarding voltage is chosen so that a positive value on this axis corresponds to a negative physical voltage on the grid (and thus a rejection of lower energy electrons).  In this example, the signal is fairly broad across all five collectors, indicating that the cloud density is not strongly peaked around the beam.  It also falls off quickly with retarding voltage, indicating that the majority of cloud particles have low energy.  The RFA signal is expressed in terms of current density in nA/mm$^2$, normalized to the transparency of the RFA beam pipe and grids.  In principle, this gives the time averaged electron current density incident on the beam pipe wall.  The beam conditions are given as ``1x45x1.25~mA e$^+$, 14~ns, 5.3~GeV."  This notation, which will be used throughout this section, indicates one train of 45 bunches, with 1.25~mA/bunch (1~mA = $1.6\times10^{10}$ particles), with positrons, 14~ns spacing, and at beam energy 5.3~GeV.

\begin{figure}
   \centering
   \begin{tabular}{c}
       \includegraphics*[width=0.45\textwidth]{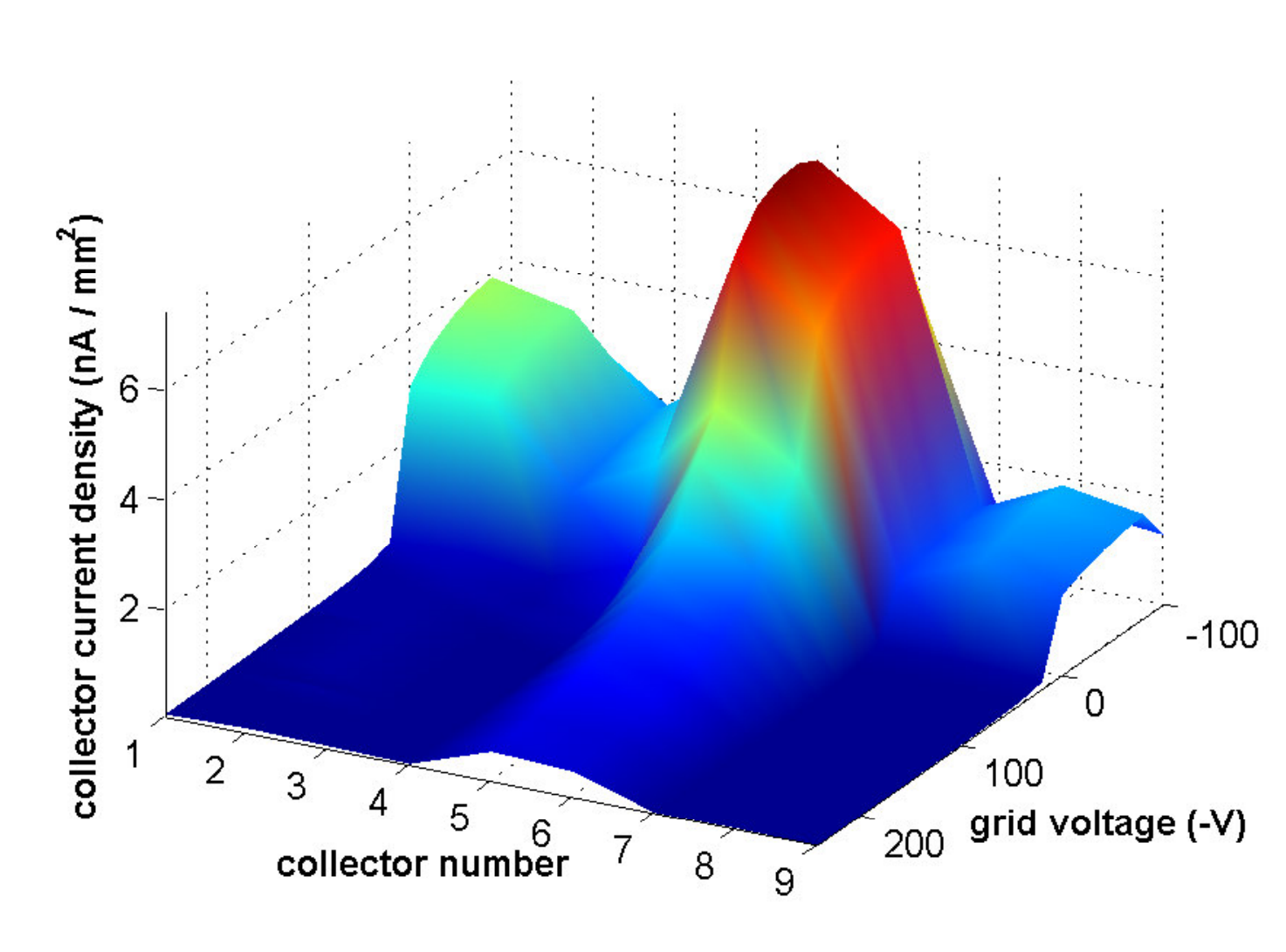} \\
       \includegraphics*[width=0.45\textwidth]{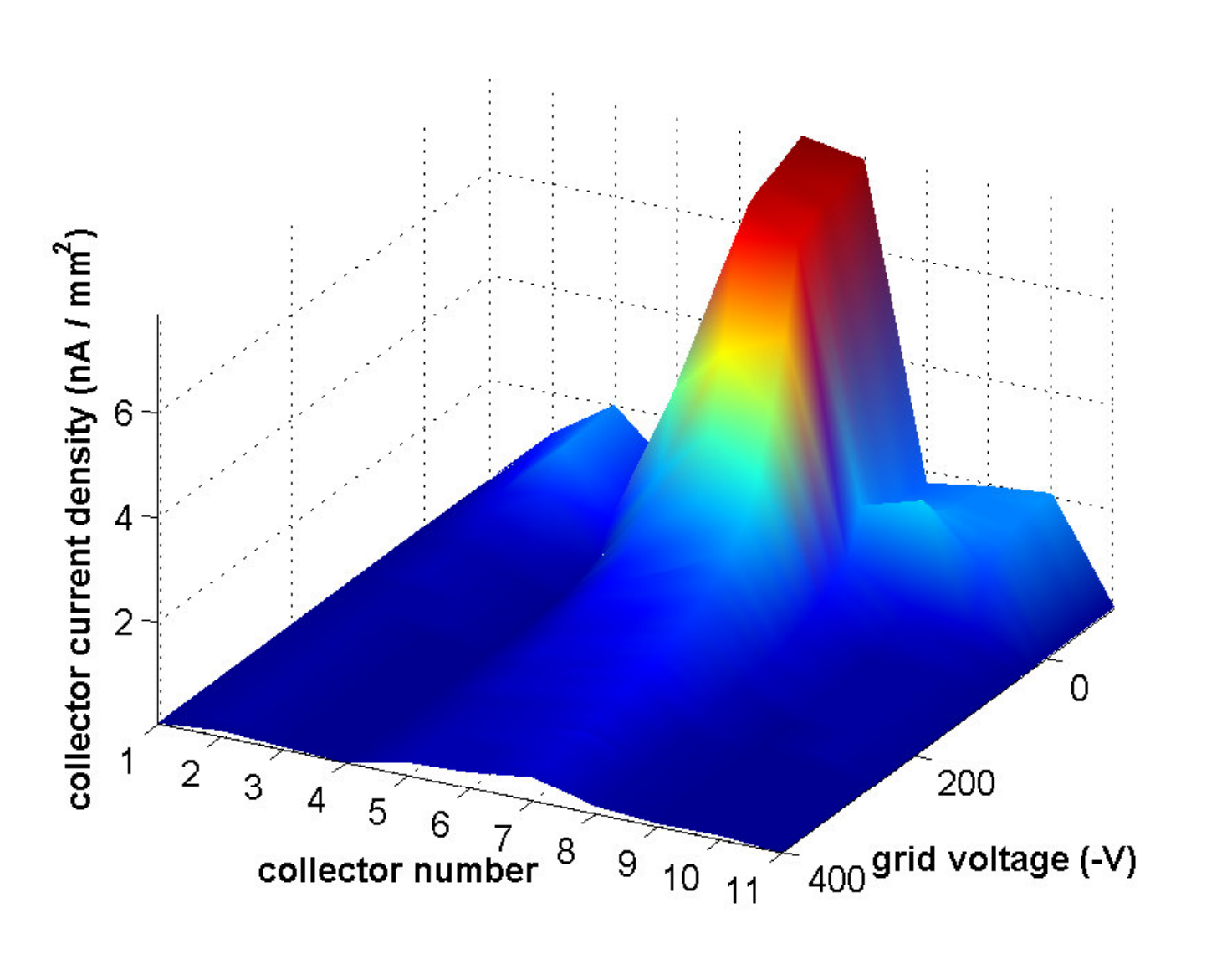}
   \end{tabular}
   \caption[Example voltage scans with thin and insertable style drift RFAs]{\label{fig:drift_example} Example voltage scans with thin (top) and Insertable Type II (bottom) style drift RFAs in the same location (Q15W). Both are TiN-coated, beam conditions are 1x45x1.25~mA, 5.3~GeV, 14~ns.}
\end{figure}

\begin{figure}
   \centering
   \includegraphics*[width=0.5\textwidth]{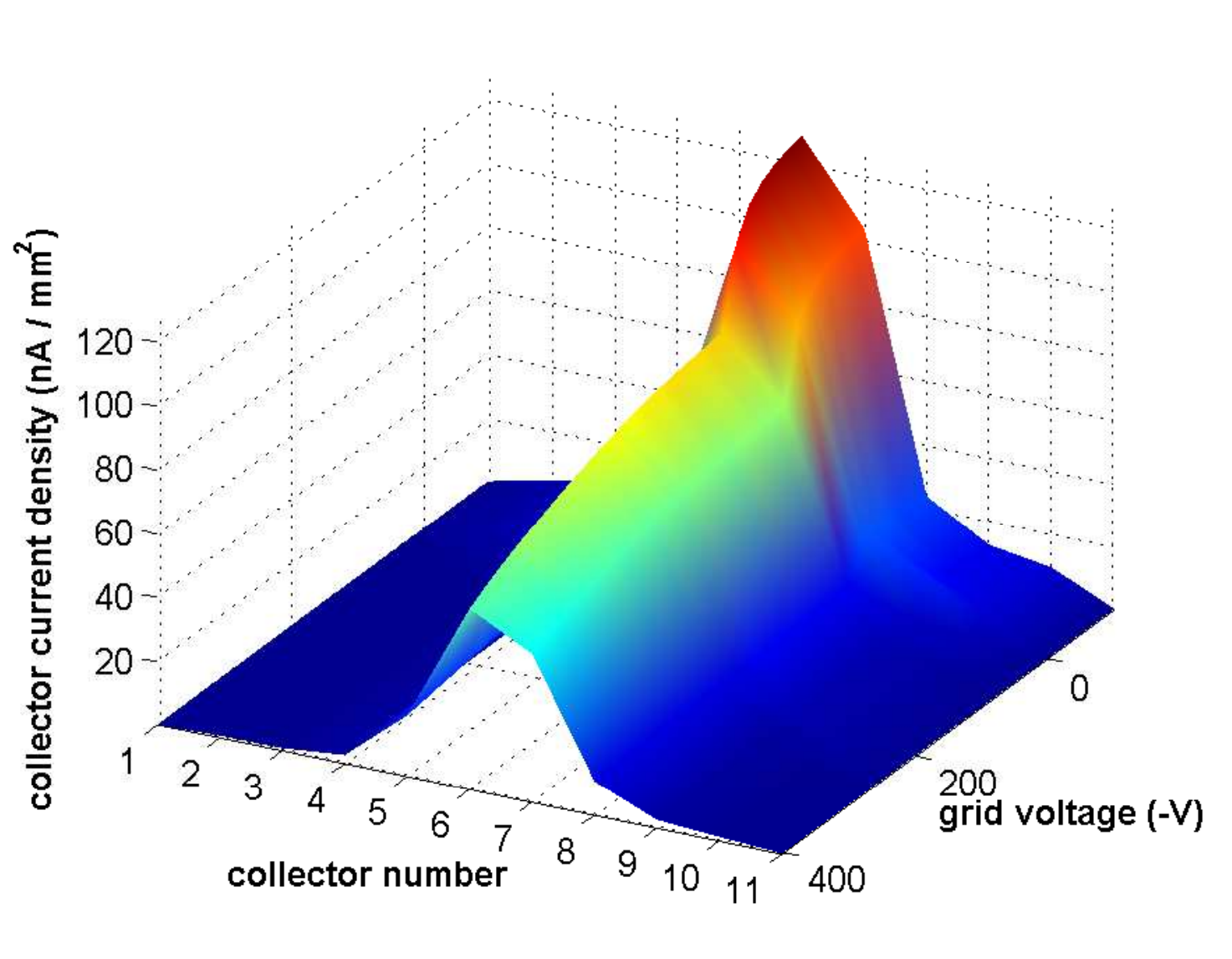}
   \caption{\label{fig:high_cur_example} Voltage scan at high bunch current, 1x20x10~mA e$^+$, 5.3~GeV, 14~ns, Insertable Type II RFA, in a TiN-coated chamber at Q15W.}
\end{figure}

As described in Section~\ref{ssec:styles}, both ``thin" and ``Insertable Type II" style RFAs have been installed at Q15E and Q15W.  Example measurements done with both of these RFA styles, in a TiN-coated chamber, can be found in Fig.~\ref{fig:drift_example}.  Again, these measurements indicate a relatively uniform, low energy cloud.  In contrast, a measurement done at high bunch current (Fig.~\ref{fig:high_cur_example}) shows a signal that is more strongly peaked in the central collector, and extends to high retarding voltage.  This implies the cloud is more concentrated in the center of the chamber, and much higher energy (due to the stronger beam kicks).

\subsection{Bunch Spacing Comparison}

\begin{figure}
   \centering
   \includegraphics*[width=0.5\textwidth]{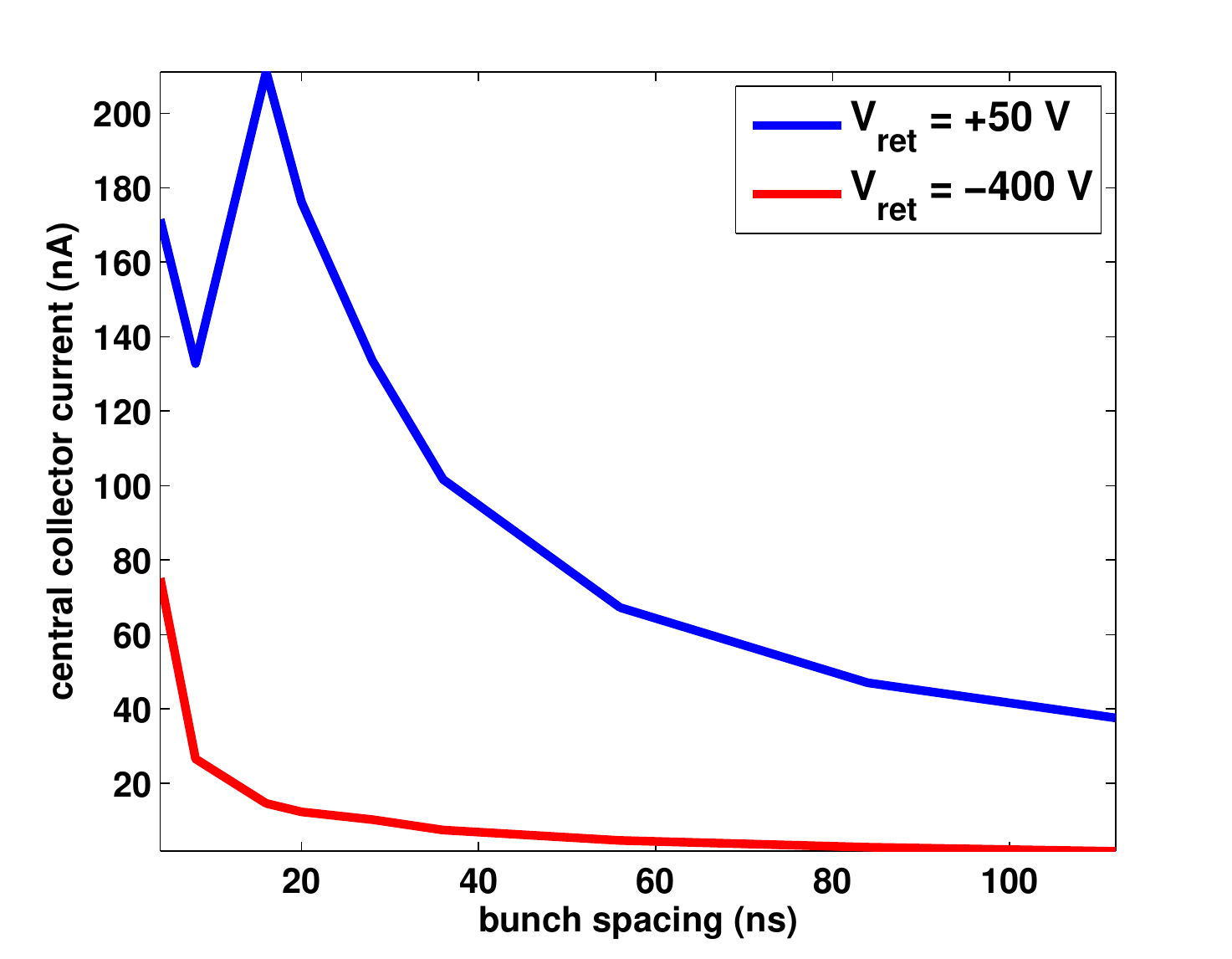}   
   \caption{\label{fig:spacing_comparison} Central collector signal as a function of bunch spacing, 1x20x3.6~mA e$^+$, 5.3~GeV, in an aC coated chamber at Q15W.}
\end{figure}

Although our RFA measurements are not time resolved, we can probe the time scale of cloud development by examining the RFA response as a function of bunch spacing, which can be varied in 4~ns increments.  Fig.~\ref{fig:spacing_comparison} shows such a comparison for the Q15W aC coated RFA.  We observe that the signal at high retarding voltage (i.e. the number of high energy cloud particles) is highest for the 4~ns data, and falls off quickly and monotonically with increasing bunch spacing.  With short bunch spacing, a typical electron will receive multiple beam kicks before colliding with the vacuum chamber, gaining 100s of eV in the process.  However, the total signal (including high and low energy electrons) is actually highest for 16~ns.  This is consistent with a multipacting resonance~\cite{HEACC77:GROBNER,PAC03:RPPG002}, in which the kick from the beam gives secondary electrons near the vacuum chamber wall just enough energy to reach the opposite wall in time for the next bunch.  These electrons generate more secondaries, which are again given energy by the beam.  This process continues, resulting in a resonant buildup of the cloud.  The resonant condition is given by Eq.~\ref{eq:grob}.  Here $t_b$ is the bunch spacing, $b$ is the chamber half-height, $r_e$ is the classical electron radius, and $N_b$ is the bunch population.  For the beam conditions in Fig.~\ref{fig:spacing_comparison}, this comes out to 13~ns, consistent with the 16~ns peak observed.

\begin{equation}
    \label{eq:grob}
    t_b = \frac{b^2}{c r_e N_b}
\end{equation}

\subsection{Mitigation Comparisons}

An important component of the \cesrta~program is the direct comparison of different electron cloud mitigating coatings, tested at Q14E, Q15E/W, and L3.  In this section we compare ``current scans" (RFA signal as a function of beam current), for different mitigations in each of the instrumented sections of CESR.  The transverse distribution observed at a given beam current was substantially the same for different chambers, so the plots shown below average over the RFA collectors.  All of the measurements were done with the retarding grid biased to +50~V in order to measure cloud electrons of all energies.

In general, current scans show a characteristic behavior as a function of beam current.  At low current ($\lesssim$ 40~mA), the cloud is dominated by primary photoelectrons, and the RFA signal increases linearly with current.  At higher currents, the increased presence of secondary electrons leads to a nonlinear increase in signal.  At very high currents, the cloud growth is limited by space charge.

\subsubsection{Comparison of adjacent chambers at Q14E}

\begin{figure}
   \centering
   \includegraphics*[width=0.5\textwidth]{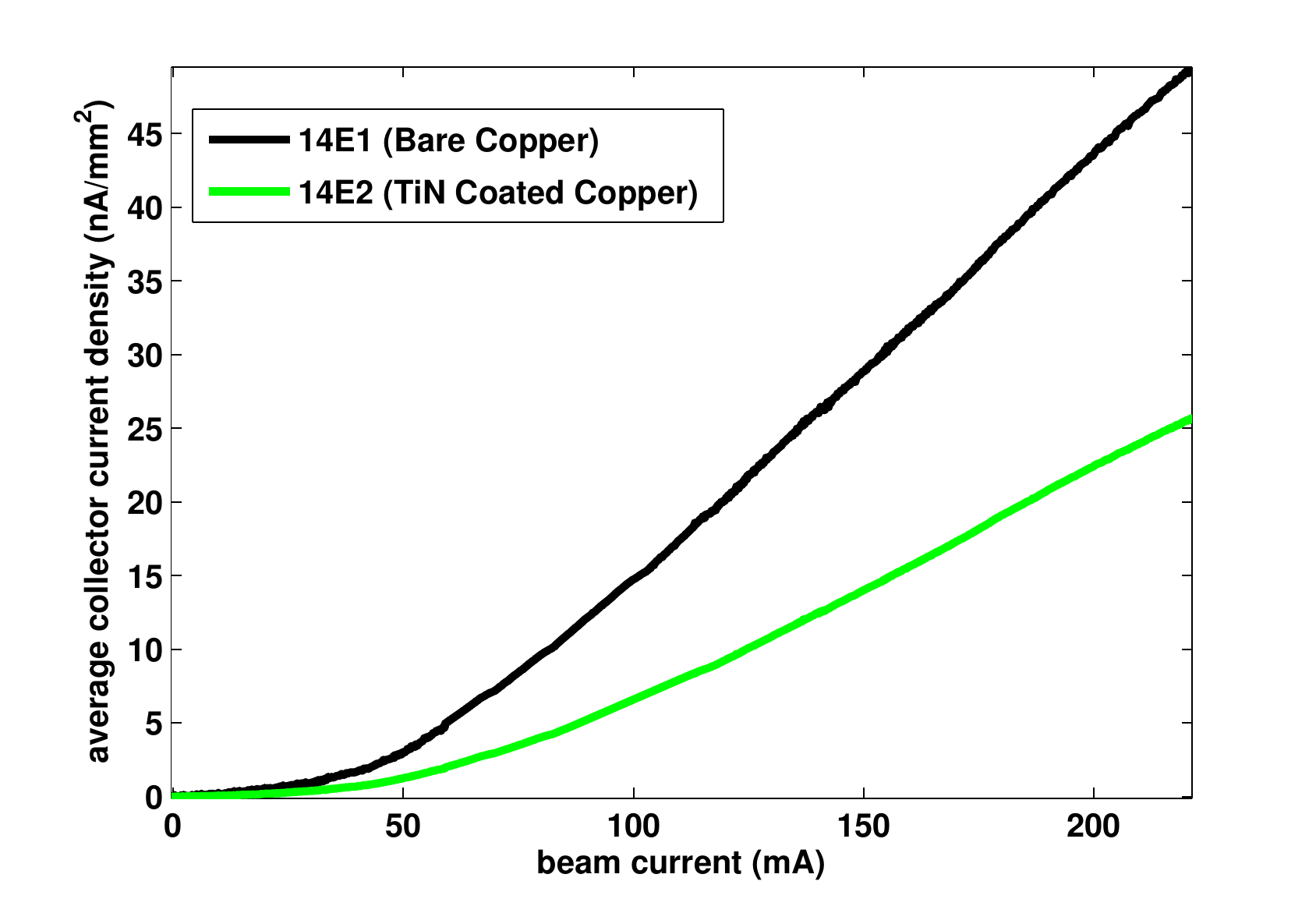}
   \caption[Comparison of insertable drift RFAs]{\label{fig:seg_compare} Comparison of insertable drift RFAs, 1x20 e$^+$, 5.3~GeV, 14~ns}
\end{figure}

Fig.~\ref{fig:seg_compare} compares a current scan measurement done simultaneously with two adjacent RFAs in the Q14E section (Section~\ref{ssec:14EW}), one in a bare copper chamber, and one in a TiN-coated copper chamber.  Here we compare the average collector current density in the two detectors, as a function of beam current, and find that it is lower in the coated chamber by a factor of two.  The photon flux is actually about 50\% higher in the TiN coated chamber, so a more direct comparison would show an even larger improvement.

\subsubsection{\label{sssec:meas_15EW} Comparisons of chambers with different coatings installed at the same locations at Q15E/W}

\begin{figure}
\centering
\begin{tabular}{c}
\includegraphics[width=.45\textwidth]{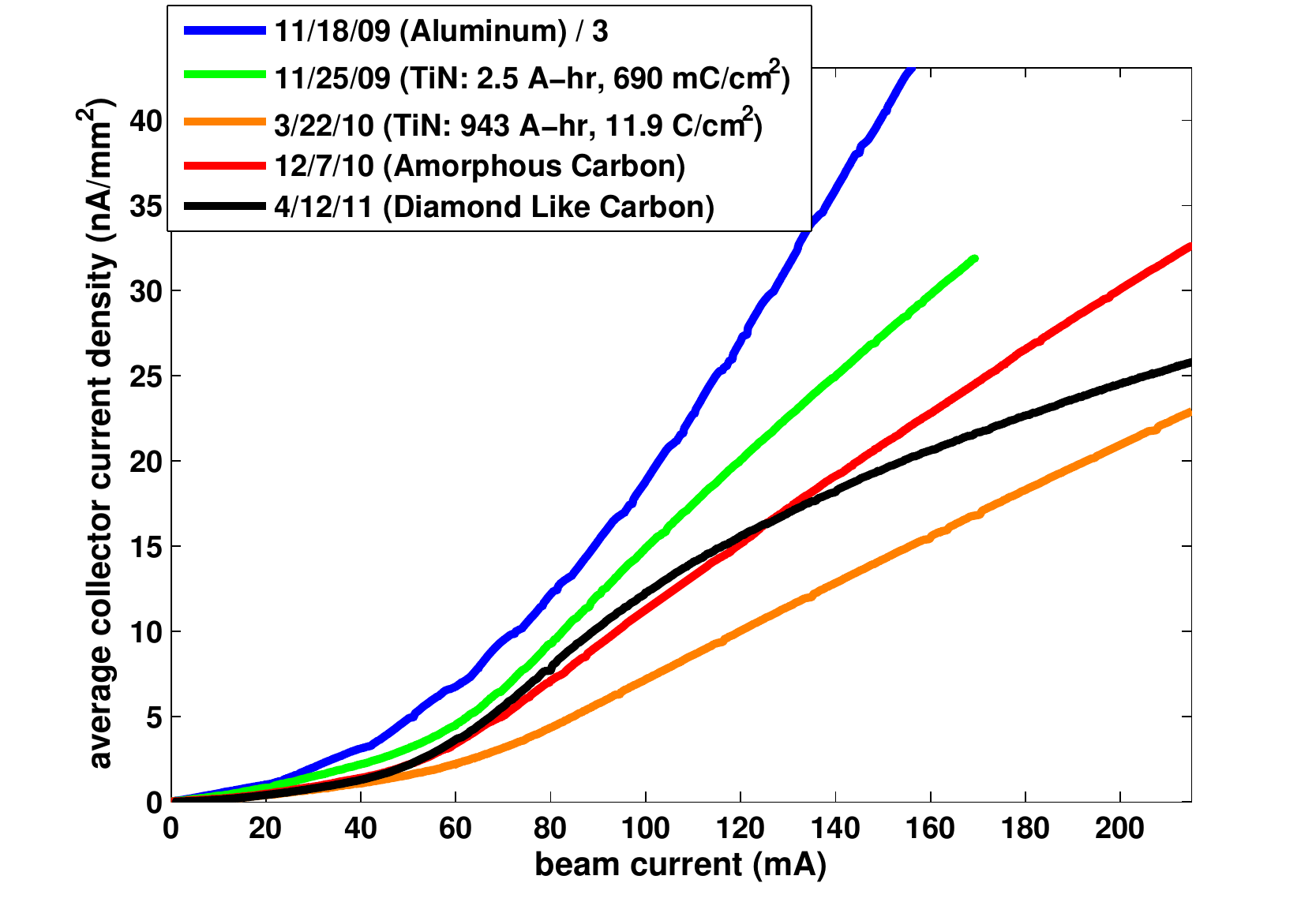} \\ 
\includegraphics[width=.45\textwidth]{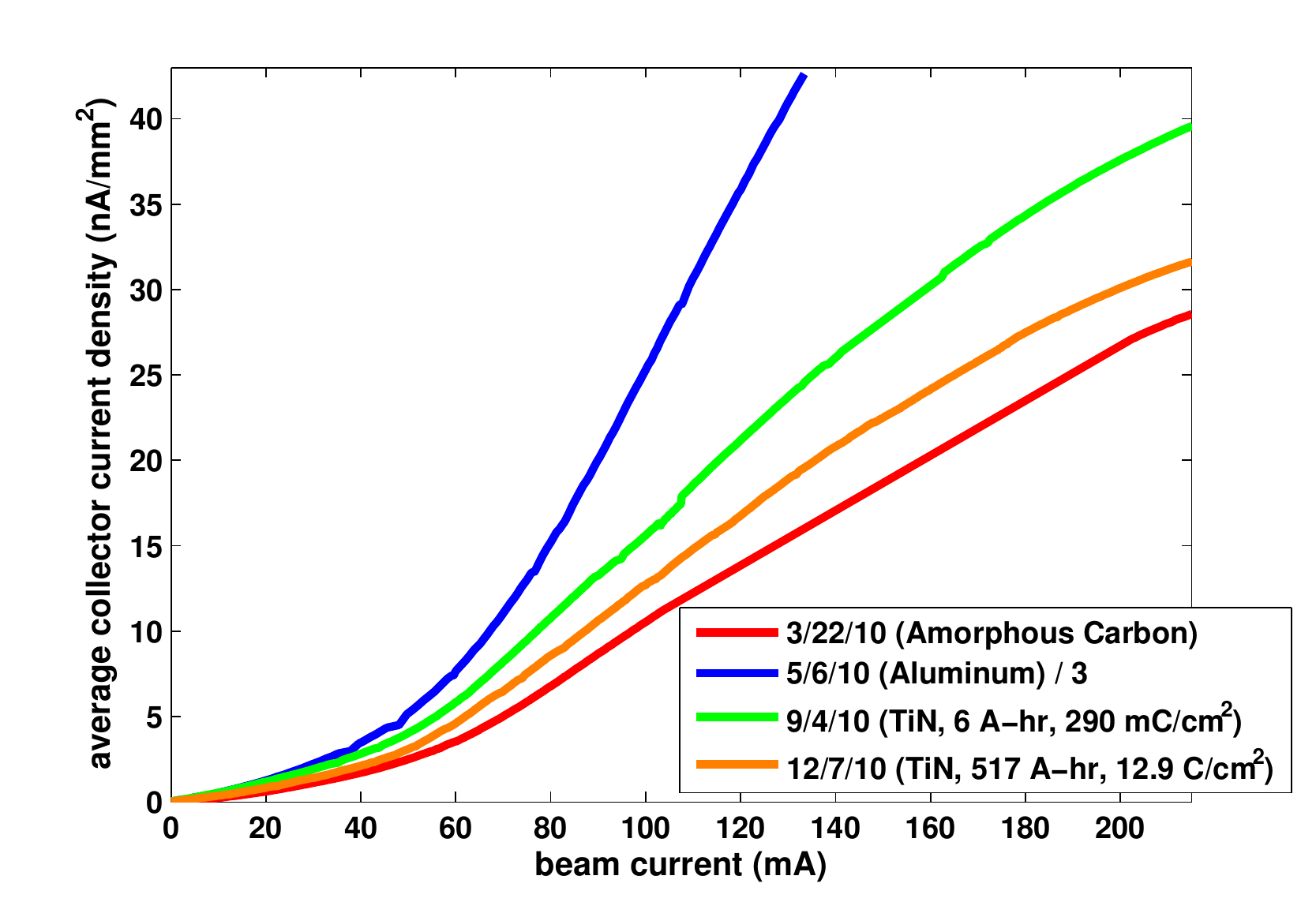} \\ 
\end{tabular}
\caption[Comparison of different beam pipe coatings for drift RFAs: 20 bunches of positrons, 14ns spacing, 5.3GeV.]{\label{fig:drift_compare_pos} Comparison of different beam pipe coatings, Q15E (top), and Q15W (bottom) drift RFAs.  Plots show average collector signal vs beam current for 20 bunches of positrons with 14~ns spacing, at beam energy 5.3~GeV.  Note that the aluminum chamber signals are divided by 3.  The TiN signal is plotted for two different beam doses.}
\end{figure}

\begin{figure}
\centering
\begin{tabular}{c}
\includegraphics[width=.45\textwidth]{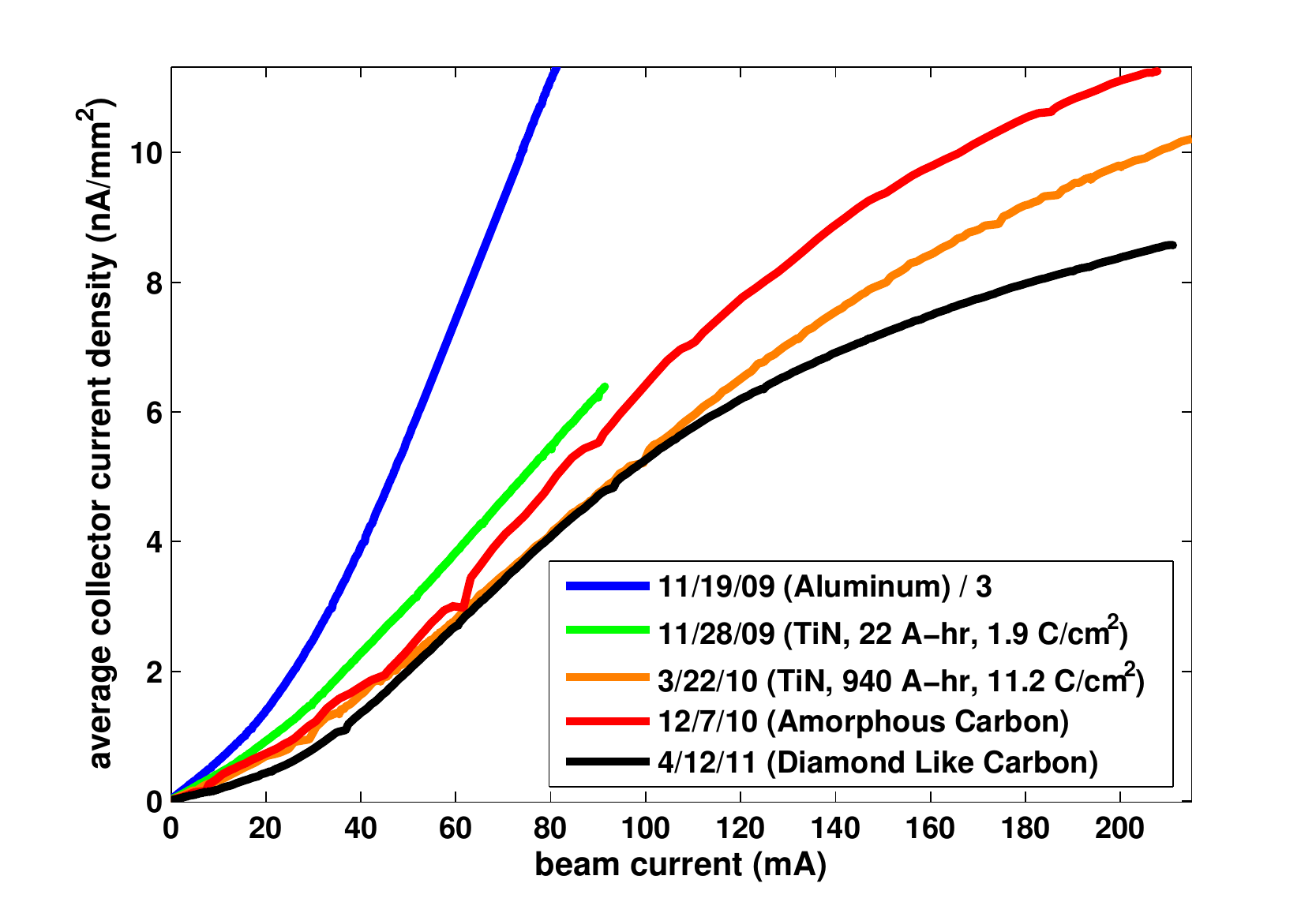} \\ 
\includegraphics[width=.45\textwidth]{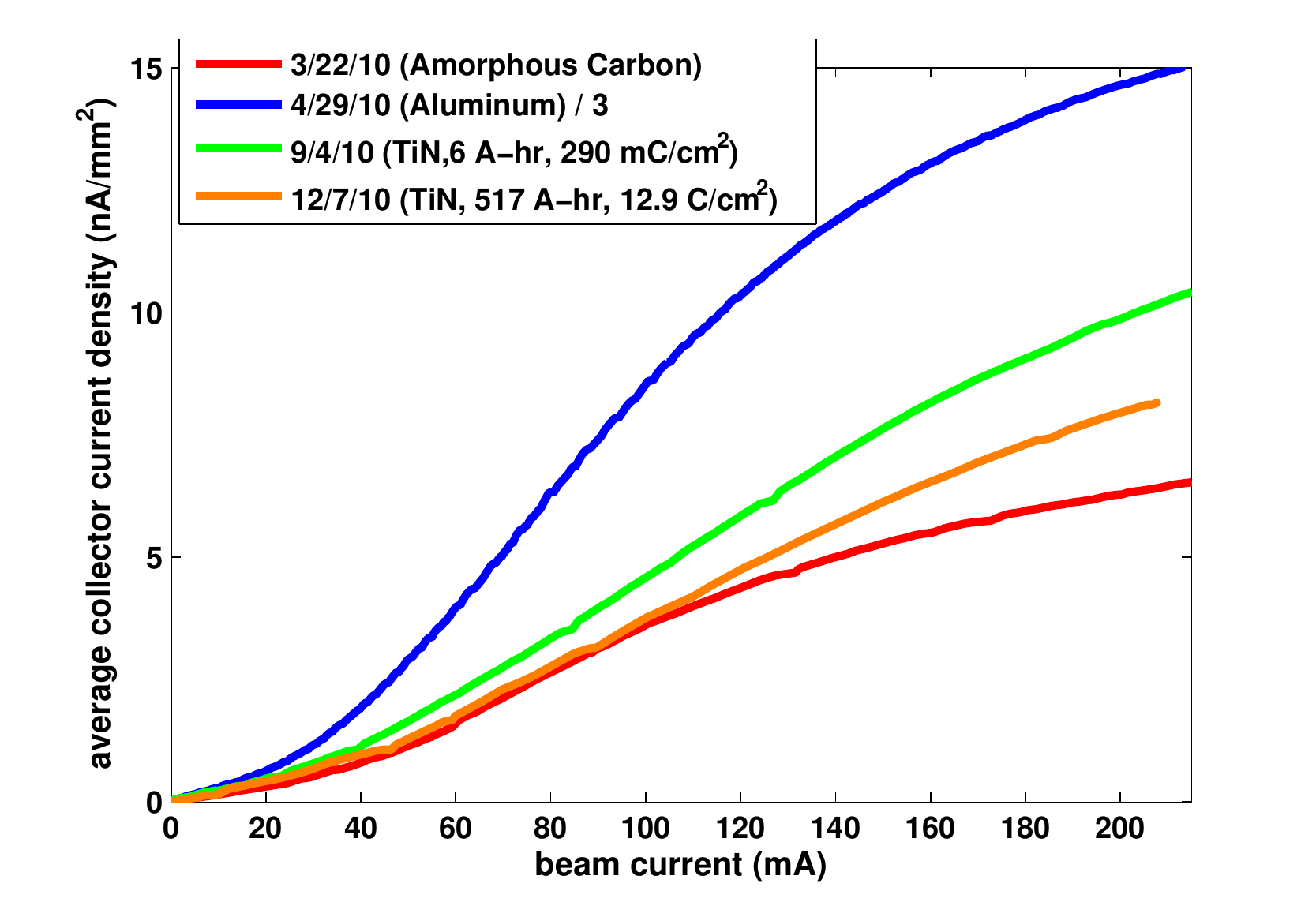} \\ 
\end{tabular}
\caption[Comparison of different beam pipe coatings for drift RFAs: 20 bunches of electrons, 14ns spacing, 5.3GeV.]{\label{fig:drift_compare_elec} Comparison of different beam pipe coatings, Q15E (top), and Q15W (bottom) drift RFAs.  Plots show average collector signal vs beam current for 20 bunches of electrons with 14~ns spacing, at beam energy 5.3~GeV.  Note that the aluminum chamber signals are divided by 3.  The TiN signal is plotted for two different beam doses.}
\end{figure}

\begin{figure}
\centering
\begin{tabular}{c}
\includegraphics[width=.45\textwidth]{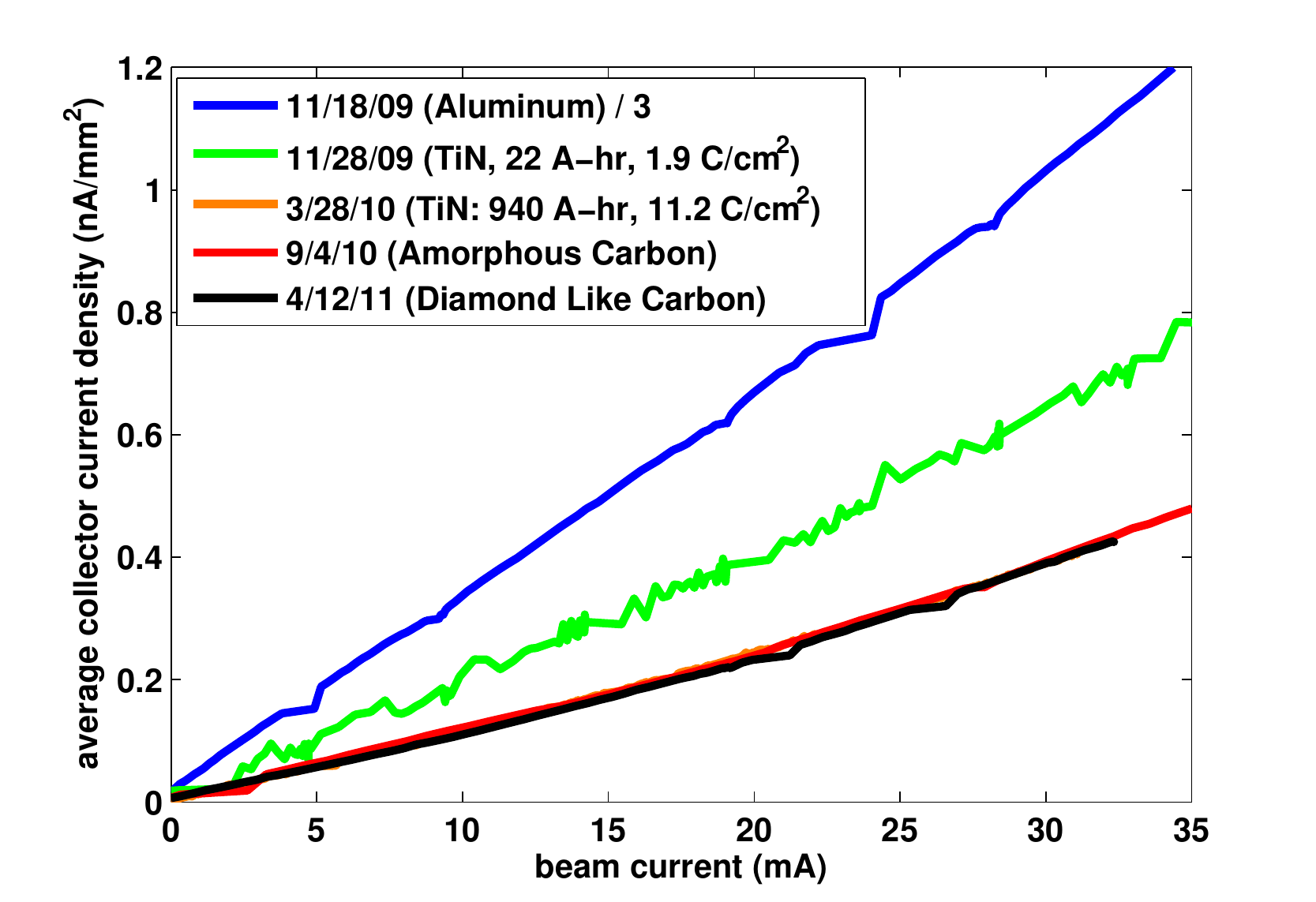} \\ 
\includegraphics[width=.45\textwidth]{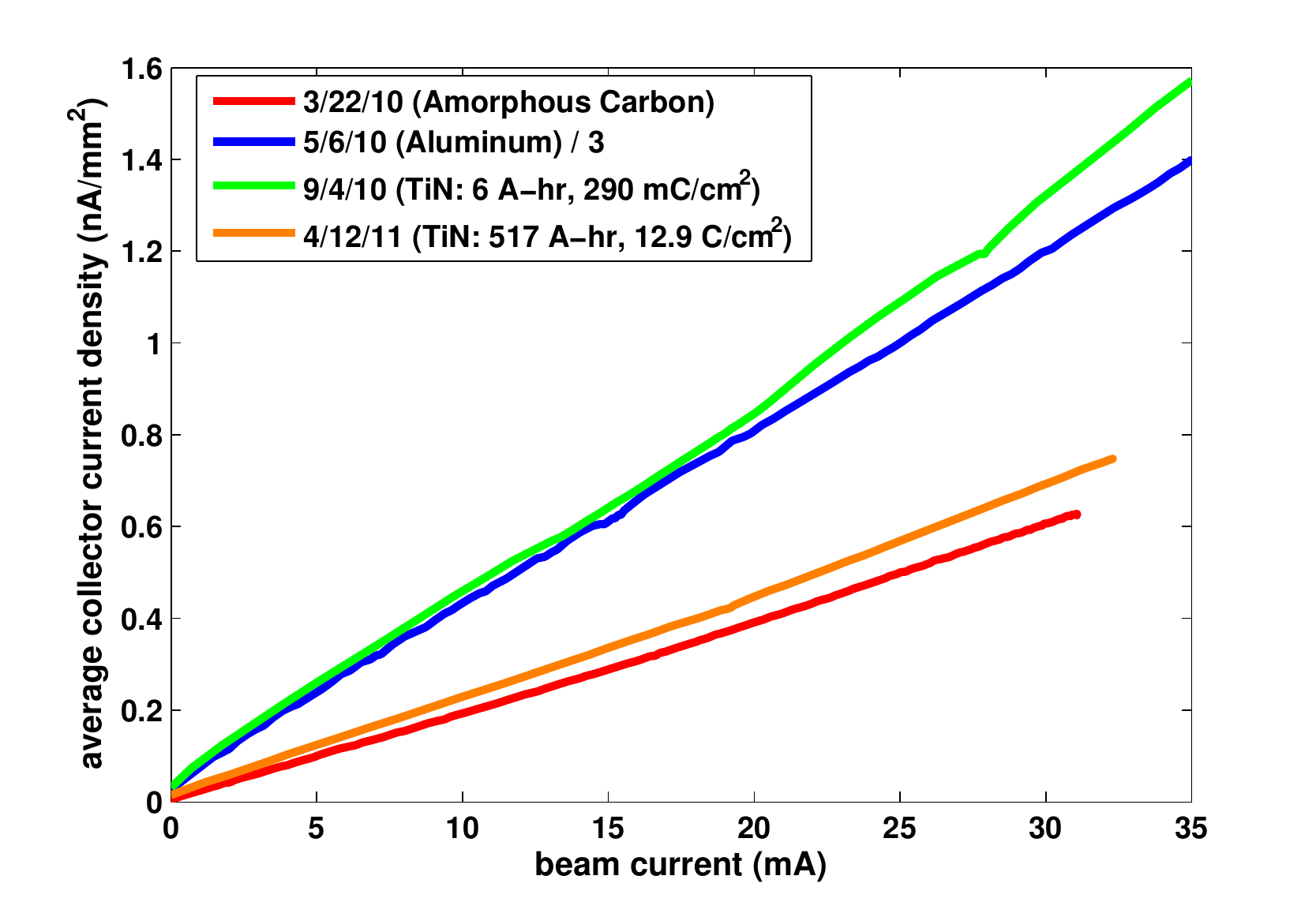} \\ 
\end{tabular}
\caption[Comparison of different beam pipe coatings for drift RFAs: 9 bunches of positrons, 280ns spacing, 5.3GeV.]{\label{fig:drift_compare_9x1} Comparison of different beam pipe coatings, Q15E (top), and Q15W (bottom) drift RFAs.  Plots show average collector signal vs beam current for 9 equally spaced (280~ns) bunches of positrons, at beam energy 5.3~GeV.  Note that the aluminum chamber signals are divided by 3.  The TiN signal is plotted for two different beam doses.  In the top plot, the curve for processed TiN is difficult to see, because it lies almost directly under the curve for aC.}
\end{figure}

The majority of our mitigation studies were done with RFAs in the Q15W and Q15E experimental sections (Section~\ref{ssec:15EW}).  The photon flux from a positron beam at Q15W is about twice that of Q15E, and vice versa for an electron beam.  Measurements have been taken at both locations with TiN and aC coatings, as well as with an uncoated aluminum chamber (see Table~\ref{tab:Q15_chamber_table}).  In addition, a chamber with DLC coating has been installed at Q15E.  By comparing measurements taken at the same location in CESR, we ensure the comparisons can be made under identical beam conditions, including photon flux.  Figs.~\ref{fig:drift_compare_pos} through~\ref{fig:drift_compare_9x1} compare the RFA signal with each of these coatings for typical sets of \cesrta~beam conditions.  The beam energy is 5.3~GeV in all cases; the comparisons are for one train of 20 bunches spaced by 14 ns (positrons in Fig.~\ref{fig:drift_compare_pos}, electrons in Fig.~\ref{fig:drift_compare_elec}) and 9 bunches of positrons spaced by 280 ns (Fig.~\ref{fig:drift_compare_9x1}).  We have generally found that data taken with 20 bunches of positrons at high current shows the biggest difference between the different chambers.  It is under these conditions that we expect to be most sensitive to the secondary electron yield, since higher bunch currents lead to higher energy electrons, which are more likely to produce secondaries.

There was some concern that these measurements could be affected by the adjacent aluminum chambers (where we expect a higher cloud density).  To address this issue, ~100 Gauss dipole magnets were installed on either side of the coated chambers, to prevent longitudinal motion of electrons into the chambers.  We found that the use of these magnets had little effect on the RFA measurements.


All coated chambers show a sizeable reduction in signal when compared to uncoated aluminum.  We have found that exposure to electron cloud bombardment significantly improved the performance of the TiN-coated chamber.  This effect, known as ``scrubbing" or ``processing," is well known~\cite{PRL109:064801}, and has been observed in direct measurements of the SEY of a TiN coated chamber~\cite{NIMA551:187to199}.  In these plots, the integrated beam current is given in A-hr.  At both Q15E and Q15W, 1 A-hr is approximately equivalent to a photon dose of $1.6 \times 10^{22}$ photons/m.  The integrated electron cloud flux (in C/cm$^2$), as measured by the RFAs, is also given.  Our measurements indicate that TiN is fully processed after an integrated electron dose of $\sim$10 C/cm$^2$ ($\sim$10$^{19}$ electrons/cm$^2$).  Conditioning of TiN coatings by electron bombardment has been measured more directly by other groups, using an electron gun to bombard samples with a known electron energy and integrated flux~\cite{NIMA469:1to12,NIMA551:187to199}. Their experiments indicate that TiN coatings are fully scrubbed with smaller bombardment doses, on the order of 0.1 - 1 C/cm$^2$.  However, simulations have indicated that even under high beam current conditions, the average energy of a cloud electron in CESR is around $\sim$50~eV, well below standard energies used for electron gun processing measurements (typically 100~eV - 1~keV).  As lower energy electrons have been shown to be less effective at processing~\cite{PhysRevSTAB.16.011002}, this may explain the discrepancy.

The aC chamber's signal was initially low, and we did not observe a significant change in signal with EC bombardment.  After extensive processing of the TiN chamber, TiN and aC showed similar mitigation performance.  Processing was also not seen in the bare Al or Cu chambers; most likely these chambers were already fully processed (due a high electron cloud density) before any RFA measurements were made.

\begin{figure}
   \centering
   \includegraphics*[width=0.5\textwidth]{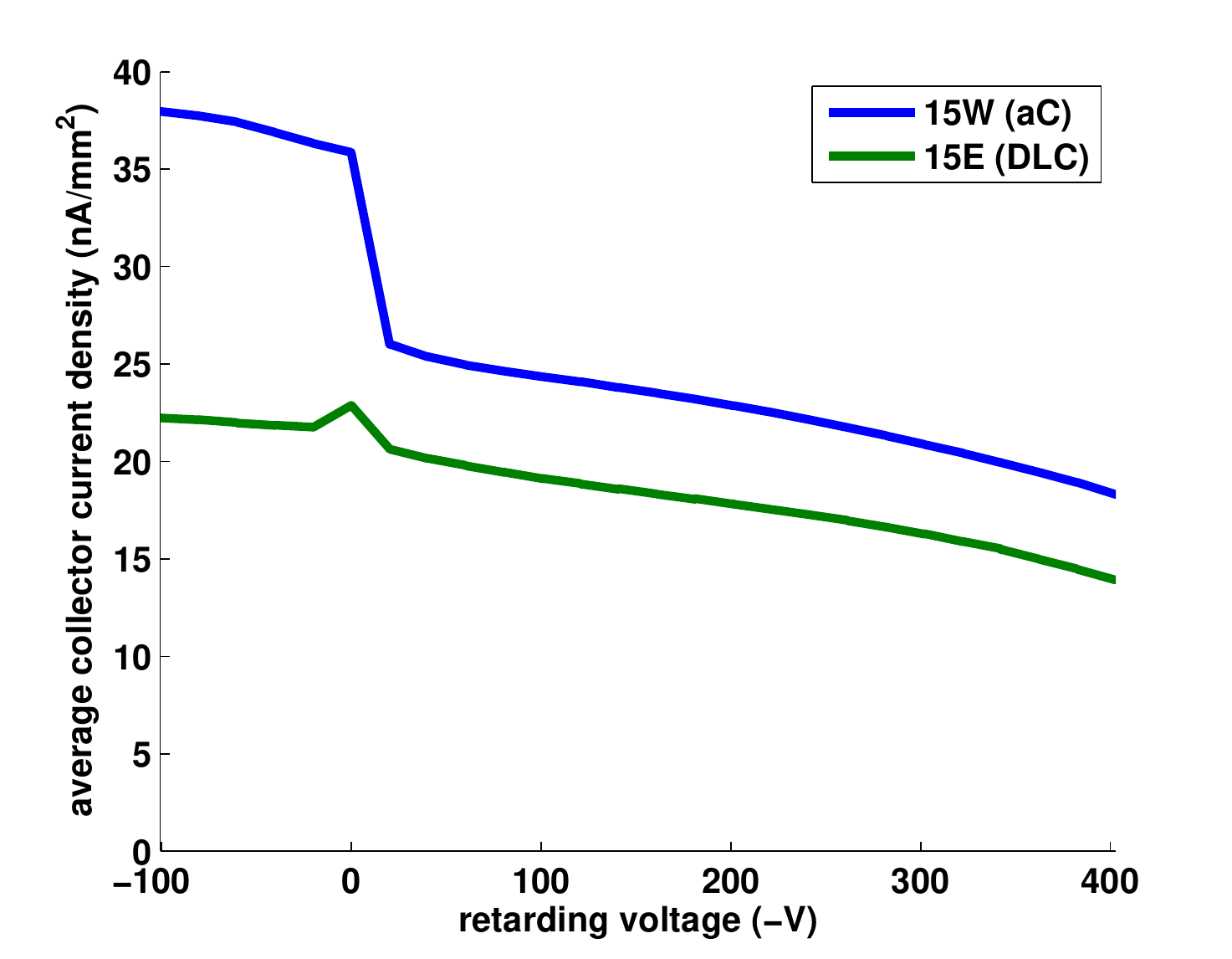}
   \caption{\label{fig:dlc_ac_comp} Comparison of amorphous and diamond-like carbon at high beam current, 1x20x10~mA e$^+$, 5.3~GeV, 14~ns.}
\end{figure}

At first glance, it appears DLC may perform better than other coatings at very high beam current.  However, it should be noted that bench measurements of the Secondary Electron Yield of DLC indicate that the material can retain charge if bombarded with a sufficiently high electron flux, thus modifying the apparent SEY~\cite{ECLOUD10:PST12}.  This effect may also be influencing the in situ measurements presented here.  Evidence for this theory can be found in Fig.~\ref{fig:dlc_ac_comp}, which compares a voltage scan done at high beam current in a DLC and aC chamber.  The aC shows an enhancement at positive retarding voltage, which is seen in almost all of our drift RFA data (see Section~\ref{ssec:bench}).  The DLC chamber instead shows a nonphysical spike at 0~V, but no enhancement at positive voltage.  This could be the result of charge around the beam pipe holes influencing the transmission of low energy electrons.

\subsubsection{\label{ssec:longterm} Long Term Behavior}

\begin{figure}
   \centering
   \includegraphics*[width=0.5\textwidth]{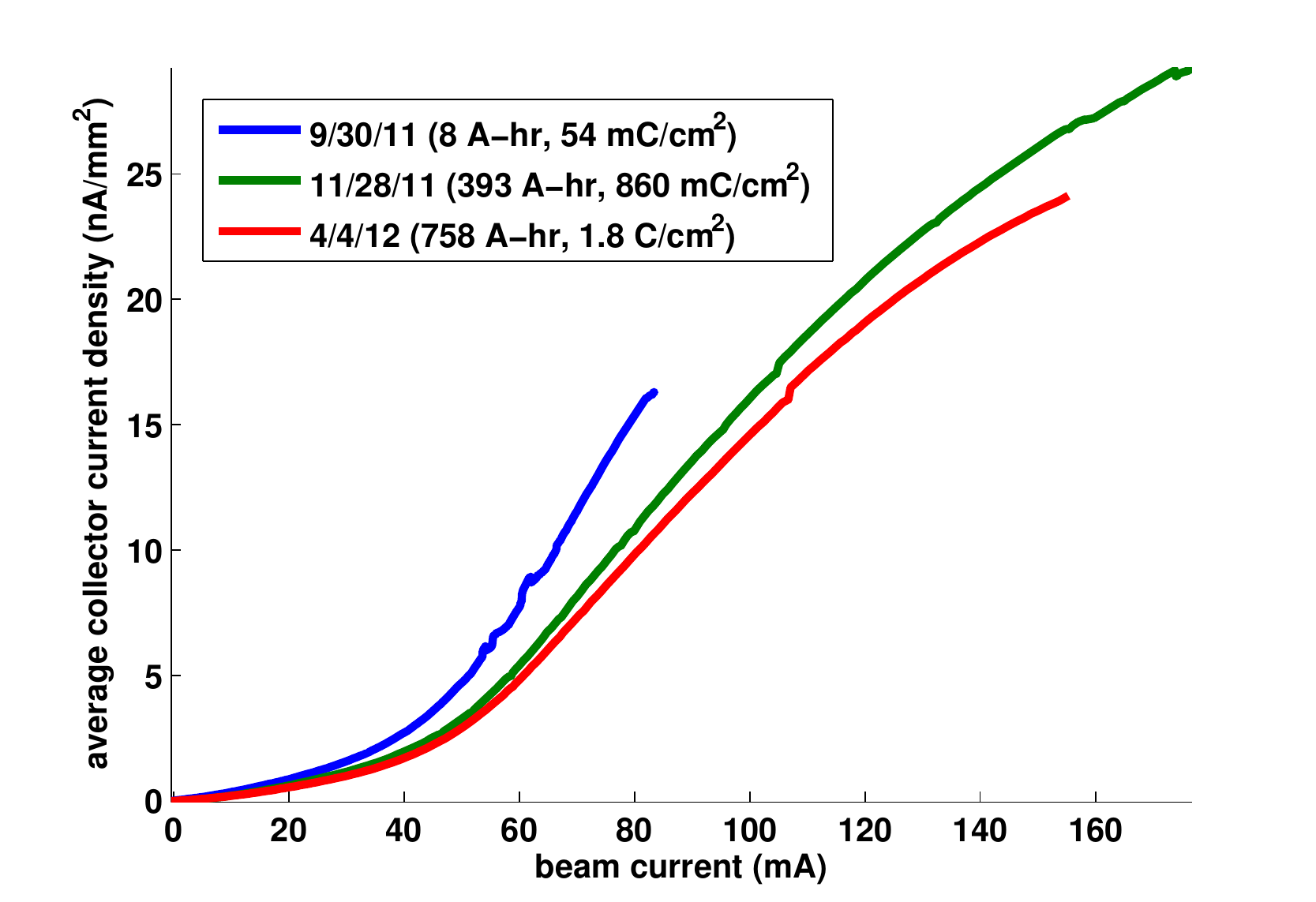}   
   \caption{\label{fig:ac2_process} Processing history in the newer Q15W aC coated chamber, 1x20 e$^+$, 5.3~GeV, 14~ns.  Integrated beam dose and electron flux is given for each measurement.}
\end{figure}

Another important issue addressed by the CESR RFA measurements is the long term reliability of various chambers and coatings.  Figures~\ref{fig:drift_compare_pos} - \ref{fig:drift_compare_9x1} show that significant processing was observed in TiN-coated chambers in both Q15E and Q15W.  In our early measurements, processing was not observed in the aC chambers.  However, more recent measurements (Fig.~\ref{fig:ac2_process}) have shown some processing in an aC coated chamber.  It is well established that the SEY of aC does not change significantly with electron cloud bombardment~\cite{PRSTAB14:071001}.  However, this decrease in signal could be explained by a reduction in the quantum efficiency, as implied by complementary measurements in a shielded pickup detector~\cite{ECLOUD12:Fri1240}.

\begin{figure}
   \centering
   \includegraphics*[width=0.5\textwidth]{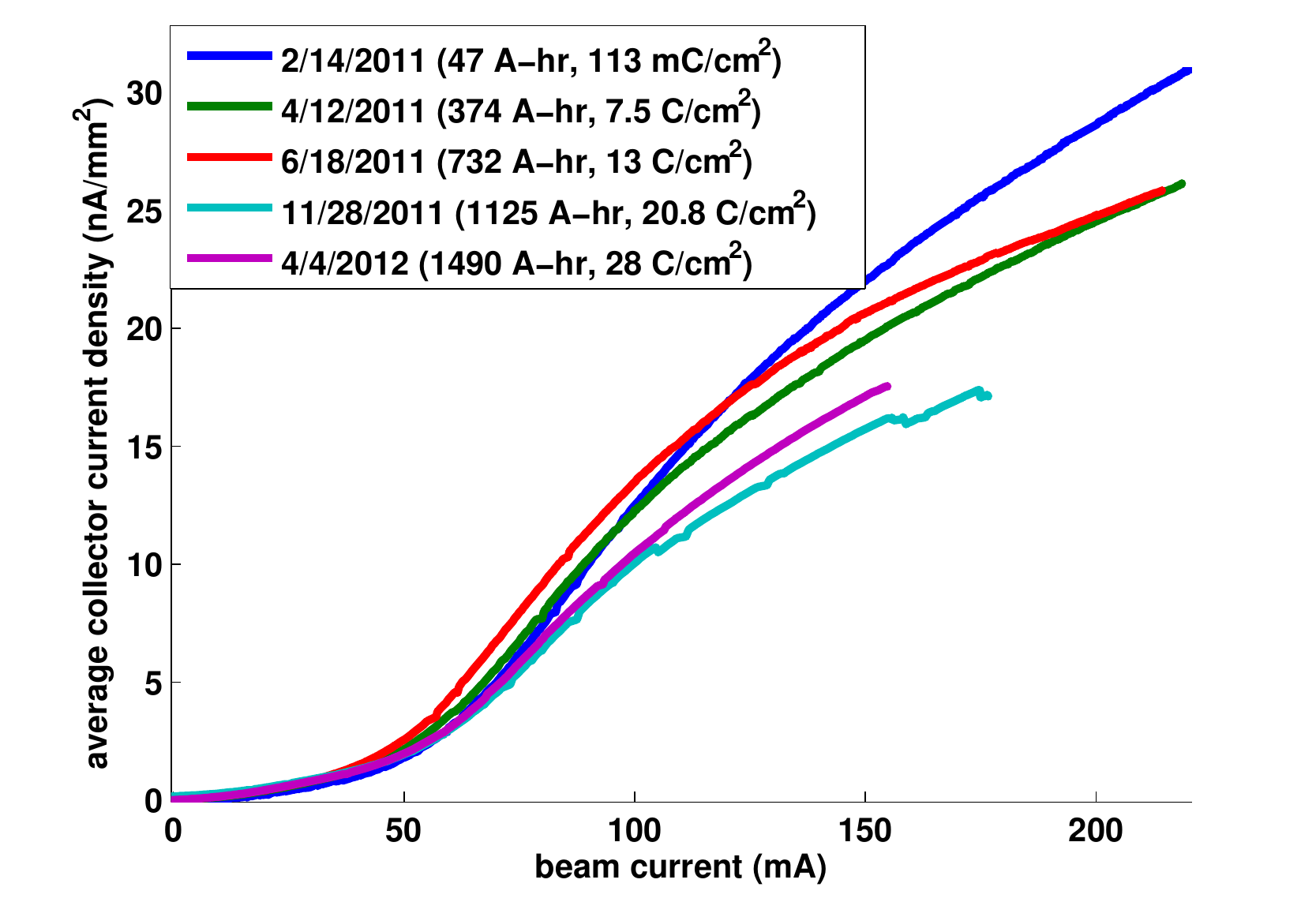}
   \caption{\label{fig:dlc_over_time} Performance of Q15E DLC chamber over time, 1x20 e$^+$, 5.3~GeV, 14~ns.  Integrated beam dose and electron flux is given for each measurement.}
\end{figure}

The signal measured in the DLC chamber varied significantly over time (Fig.~\ref{fig:dlc_over_time}).  Apart from some initial processing, the measurements in this chamber do not appear to show any obvious trend.  It is possible that the properties of the DLC depend on the recent beam history before the measurement.

\subsection{Activation and processing of NEG coated chambers in L3}

\begin{figure}
   \centering
   \includegraphics*[width=0.5\textwidth]{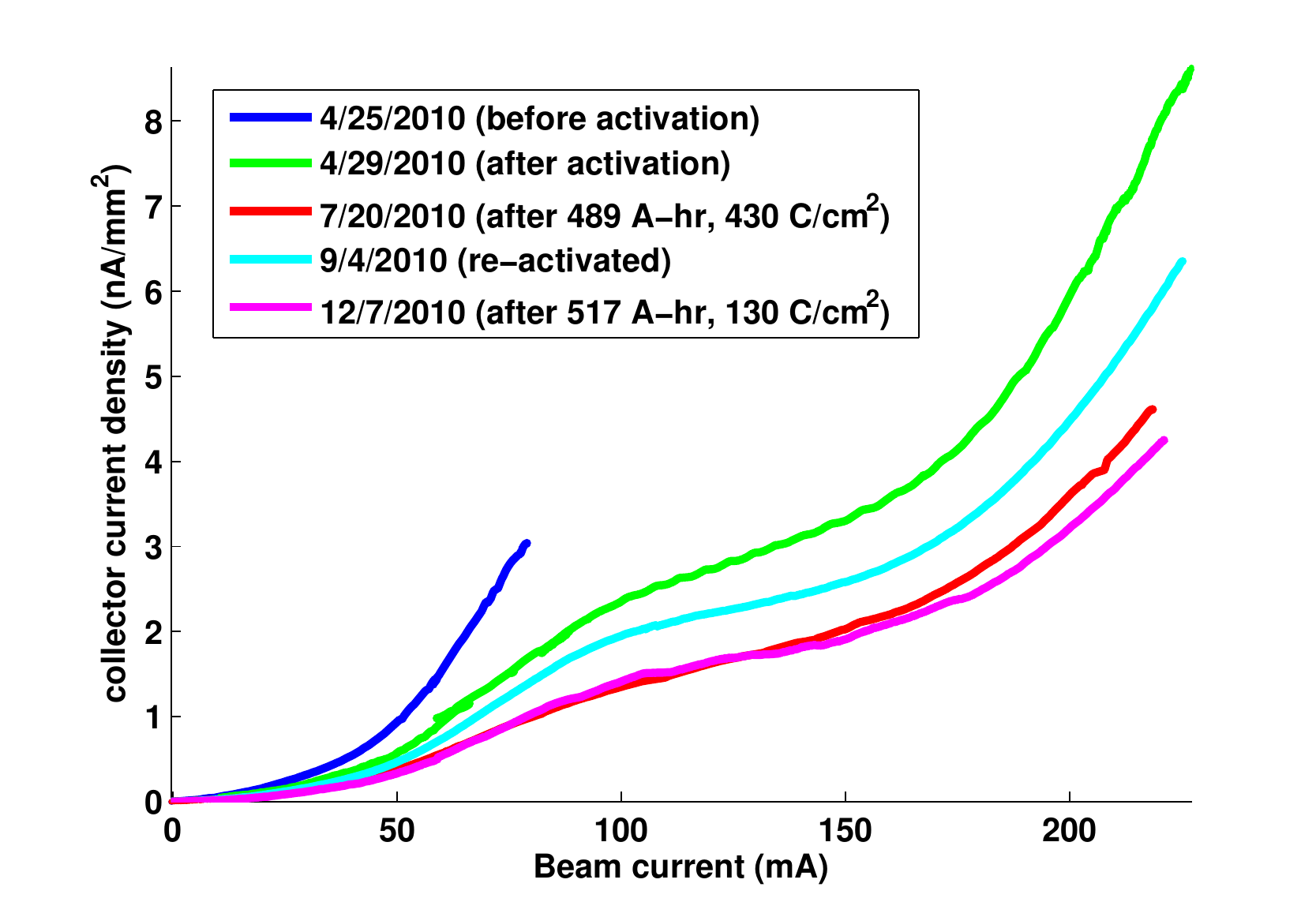}   
   \caption[NEG RFA comparison]{\label{fig:neg_rfa} NEG RFA comparison, 1x20 e$^+$, 5.3~GeV, 14~ns.}
\end{figure}

The performance of the L3 NEG coated chamber (Section~\ref{ssec:NEG}) has also been monitored using RFAs.  Fig. \ref{fig:neg_rfa} compares the current measured by one of these RFAs on several different dates, corresponding to different states of activation and processing of the NEG coating.  It was observed that both activation and initial processing reduced the signal measured by this RFA.  After a CESR down (during which the NEG was activated again), the signal rose somewhat, but it processed back down to its minimum value after a few months of beam time.  The other two detectors showed a similar trend.  These signals remained consistent in subsequent runs.

\section{Conclusions}


A major component of the \cesrta~program has been the design, commissioning, and installation of retarding field analyzers in several field-free locations in the CESR ring.  RFA designs of various types have been developed for more ideal performance or for deployment in areas with restricted space.  Particle tracking simulations were used to better quantify the relationship between the measured collector current and the cloud density, and these simulations were cross-checked via bench measurements with an electron gun.

A great deal of electron cloud data has been collected with the drift RFAs.  These data have been used to directly compare the efficacy of various electron cloud mitigating coatings (TiN, aC, DLC, and NEG).  All of the coatings tested resulted in a sizable reduction in electron cloud signal relative to an uncoated (Al or Cu) chamber.  After processing, TiN and aC showed similar mitigation performance.  DLC also appears to be effective, though there is some question about whether charging of the coating could have affected the measurement.  Also, with the possible exception of DLC, the coatings show mostly stable behavior over the long term.  More quantitative comparisons require detailed simulations, and are addressed in a separate paper~\cite{ARXIV:1402.7110}.

\section*{Acknowledgements}

The results presented in this paper were made possible by the hard work of the \cesrta~collaboration, especially L. Bartnik, M.G. Billing, J.V. Conway, J.A. Crittenden, M. Forster, S. Greenwald, X. Liu, R.E. Meller, S. Roy, D. Rubin, S. Santos, R.M. Schwartz, J. Sikora, K. Sonnad, and C.R. Strohman.  We are also grateful to S. Calatroni and G. Rumolo at CERN for providing us with the aC coated chambers, and S. Kato at KEK for the DLC chamber.

This work was supported by the US National Science Foundation (PHY-0734867 and PHY-1002467) and Department of Energy (DE-FC02-08ER41538).

\bibliographystyle{medium}
\bibliography{rfa_drift_nim}

\newpage

\appendix

\section{\label{sec:electronics} Electronics}

\begin{figure*}
  \begin{minipage}{\textwidth}
  \centering
  \includegraphics[width=.9\textwidth]{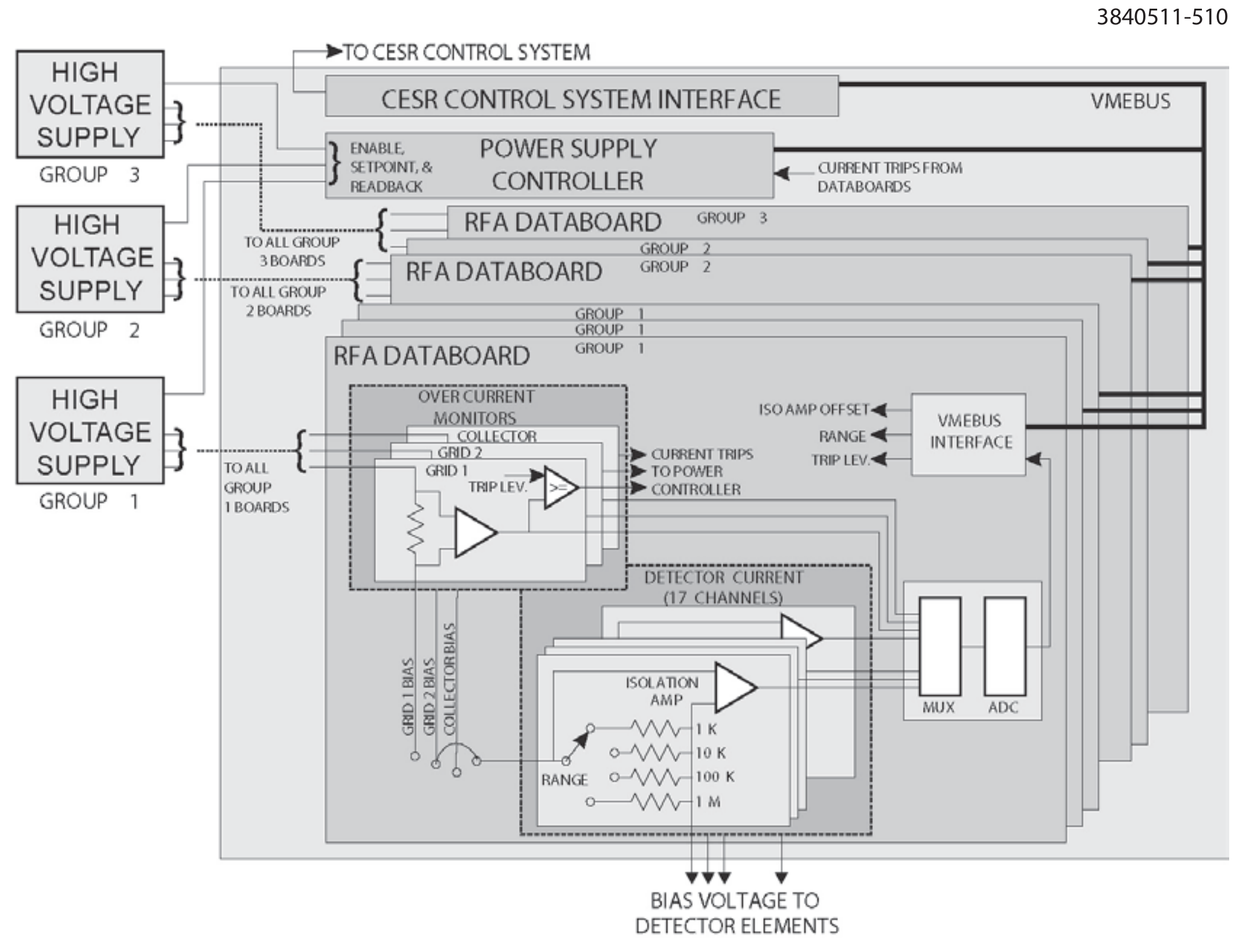}
  \caption[Schematic showing the high voltage power supply system and the RFA current monitor boards]{\label{fig:APS_RFA_electronics} Schematic showing the high voltage power supply system and the RFA current monitor boards.}
  \end{minipage}
\end{figure*}

\begin{figure*}
  \begin{minipage}{\textwidth}
  \centering
  \includegraphics[width=.6\textwidth]{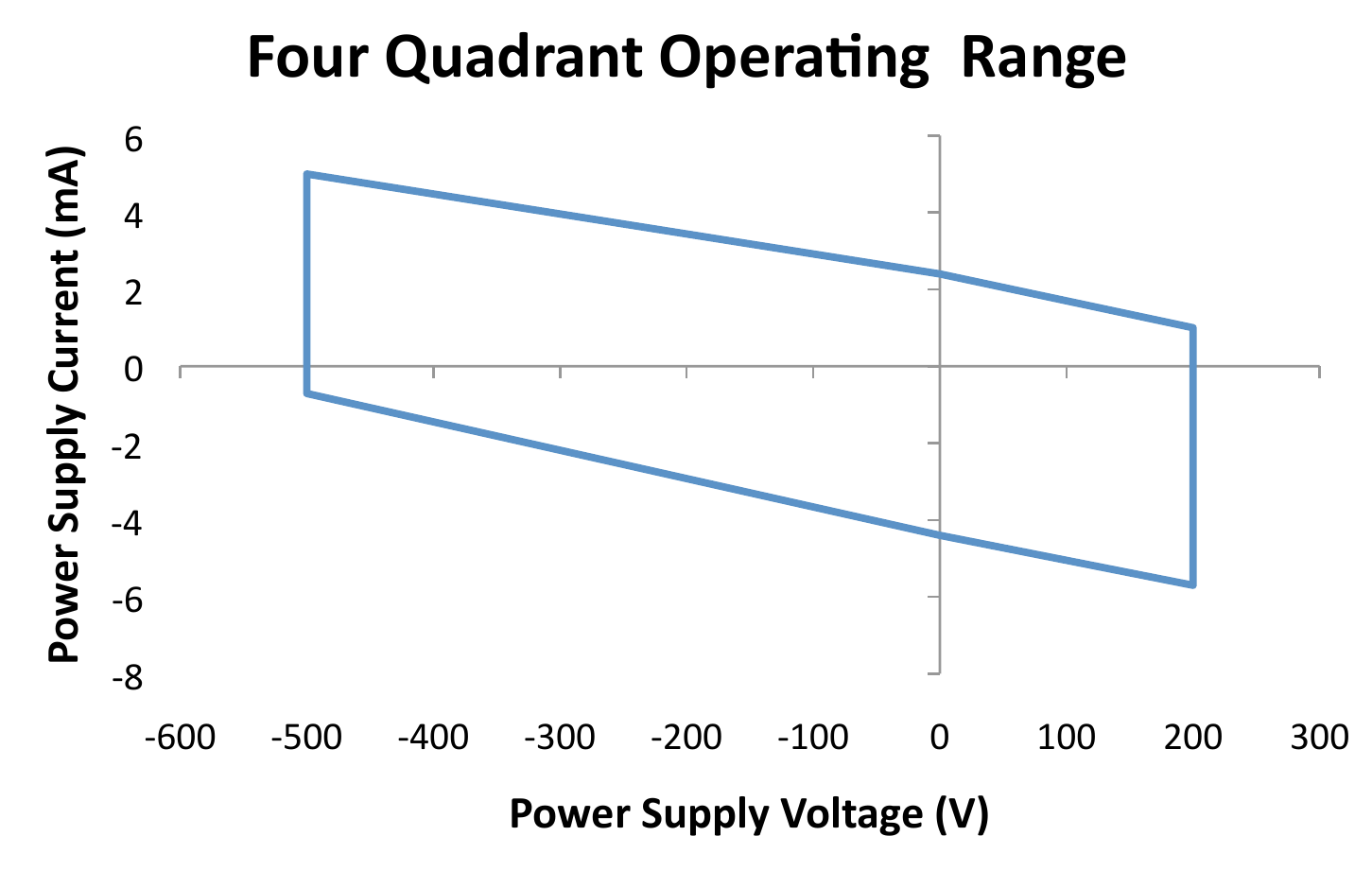}
  \caption{\label{fig:hv_supply_envelope} Operating envelope of the high voltage modules.}
  \end{minipage}
\end{figure*}

A block diagram of the RFA data acquisition system is shown in Figure~\ref{fig:APS_RFA_electronics}.  Each high voltage supply contains two four-quadrant grid supplies (described below) and a single unipolar collector supply. The standard grid supply can operate from $-500$~V to $+200$~V, with a somewhat complex operating envelope (Fig.~\ref{fig:hv_supply_envelope}); the collector supply can operate from $0$~V to $200$~V and is rated for $50$~mA.  The readout system is in a 9U VMEbus crate with a custom P3 backplane that distributes bias voltages to the databoards. This backplane is divided into three segments (which can each control up to three RFAs), each with its own HV power supply.

The four-quadrant grid supplies are implemented as a pair of EMCO F10 high voltage modules\footnote{http://www.emcohighvoltage.com/pdfs/fseries.pdf} coupled through a voltage divider.  The modules have internal voltage-scaled latch-back overcurrent protection.   This characteristic required a digital control loop to provide a programmed ramp-up sequence that will get both modules up to operating voltage without triggering an overcurrent fault.  The controller for each supply is implemented with a PIC16C773 microcontroller\footnote{http://www.microchip.com/wwwproducts/Devices.aspx?dDocName=en010176}.  The control loop starts in a coarse regulation mode where the command signals to the high voltage modules are produced by 10 bit pulse-width modulators embedded in the microcontroller, and then switches to a fine regulation mode where the commands are trimmed with 8 bit rate multipliers that are scaled to produce effective 14 bit (60~mV) resolution. The feedback is specially configured to enable high precision current measurements while the feedback loop is quiescent. Upon receipt of a voltage command, the HV control sets the voltage and allows it to stabilize. At that point, all feedback corrections are suspended for a 20 second data acquisition window. The controls for the two grid and single collector supplies in a full HV supply are configured to make this quiescent period simultaneous.

The RFA data boards distribute bias voltages to the detector elements (up to 17) and measure the current flow in each. The current is measured by an isolation amplifier looking at a series resistor (selectable as 1, 10, 100 or 1000~k$\Omega$) in the high side of the circuit with the output going to a 16-bit digitizer. The various resistors correspond to full scale ranges of 10000, 1000, 100, and 10~nA. The finest resolution is 0.15~pA.


\end{document}